\documentstyle[12pt,draft,aas_macros]{nature}
\font\large=cmbx10 scaled \magstep3

\newcommand{\be}{\begin{equation}}
\newcommand{\ee}{\end{equation}}

\newcommand{\ifm}[1]{\relax\ifmmode#1\else$\mathsurround=0pt #1$\fi}
\newcommand{\kms}{\ifmmode\,{\rm km}\,{\rm s}^{-1}\else km$\,$s$^{-1}$\fi}
\newcommand{\kpc}{\ifmmode\,{\rm kpc}\else kpc\fi}
\newcommand{\Mpc}{\ifmmode\,{\rm Mpc}\else kpc\fi}

\newcommand{\ltsima}{$\; \buildrel < \over \sim \;$}
\newcommand{\lsim}{\lower.5ex\hbox{\ltsima}}
\newcommand{\gtsima}{$\; \buildrel > \over \sim \;$}
\newcommand{\gsim}{\lower.5ex\hbox{\gtsima}}

\def\LCDM{$\Lambda$CDM}

\begin{document}

\title{\Large \bf Early gas stripping as the origin of the darkest galaxies in the Universe}
\author{
  L. Mayer$^{1,2}$,
  S. Kazantzidis$^{3,4}$,
  C. Mastropietro$^5$, 
  \& J. Wadsley$^6$
\institute{$^1$Institut f\"ur Astronomie, ETH Z\"urich, Wolfgang-Pauli-Strasse 16  , 
CH-8093 Z\"urich, Switzerland. \\
$^2$Institute for Theoretical Physics, University of Zurich, 
Winterthurestrasse 190, CH-8057 Z\"urich.\\ 
$^3$Kavli Institute for Particle Astrophysics and Cosmology,
   Department of Physics, Stanford University, P.O. Box 20450, MS 29, Stanford, CA 94309 USA.\\
$^4$Kavli Institute for Cosmological Physics,
   Department of Astronomy \& Astrophysics,
   The University of Chicago, Chicago, IL 60637 USA.\\
$^5$Universit\"ats Sternwarte M\"unchen, Scheinerstrasse 1, D-81679 M\"unchen , Germany.\\
$^6$Department of Physics and Astronomy, McMaster University, Hamilton, ON L8S 4M1, Canada.
}
}   

\date{\today}{}
\headertitle{The origin of the darkest galaxies in the Universe}
\mainauthor{Mayer et al.}

\summary{

The known galaxies most dominated by dark matter (Draco, Ursa Minor and Andromeda IX)
are satellites of the Milky Way and the 
Andromeda galaxies$^{1-4}$. They are members of a class of faint galaxies,
devoid of gas, known as dwarf spheroidals$^{3-5}$, and have by far
the highest ratio of dark to luminous matter$^{3,6}$. 
None of the models proposed to unravel their origin$^{7-10}$
can simultaneously explain their exceptional dark matter content and
their proximity to a much larger galaxy.
Here we report simulations showing that the progenitors of these galaxies were probably gas-dominated 
dwarf galaxies that became satellites of a larger galaxy earlier than the other dwarf spheroidals. 
We find that a combination of tidal shocks and ram pressure swept away the entire gas content of 
such progenitors about ten billion years ago because heating by the cosmic ultraviolet background 
kept the gas loosely bound: a tiny stellar 
component embedded in a relatively massive dark halo survived until today. All luminous galaxies should 
be surrounded by a few extremely dark-matter-dominated dwarf spheroidal satellites, and these should have the shortest 
orbital periods among dwarf spheroidals because they were accreted early.}

\maketitle


Draco, Ursa Minor and Andromeda IX have mass-to-light ratios ($M/L$) larger than $100$,
but the majority of the other dwarf spheroidals (dSphs) in the Local Group have a lower $M/L$, of order 10-30$^1$, 
typical among dwarf galaxies$^{11, 12}$. Another important difference is that Draco and Ursa Minor nearly stopped forming stars more than 10 billion years ago,
while other dSphs continued to form stars for many billions of years$^2$.
The modest potential well of these extreme dwarfs cannot be the one property that determined their nature.
Their halo masses are too large to invoke suppression of gas accretion 
owing to the cosmic ultraviolet background at high redshift$^{9,13}$ 
or blow-out due to supernovae winds$^{14}$.
Tidal shocks occurring as a dwarf repeatedly approaches the primary galaxy can transform rotationally supported
systems resembling dwarf irregular galaxies (dIrrs) into systems dominated by random motions, similar 
to dSphs$^{10}$. This tidal stirring can explain why dSphs are more clustered around the primary
galaxies relative to dIrrs but it leaves a significant gas component inside the dwarf, so that
star formation can continue for several billions of years instead of being truncated early.
Ram pressure in a hot gaseous corona could strip their gas completely$^{7,15}$ but the limitations of simulations 
so far have not allowed for firm predictions. For instance, existing calculations keep the 
structure of the stars and halo fixed in time and neglect
radiative cooling and heating of the gas$^{16,17}$. 

These earlier studies have explored the effect of a single gas removal mechanism and are, 
at best, only  qualitatively consistent with the current structure formation paradigm, a model with 
cold dark matter and a cosmological constant  ({\LCDM}). Recent attempts to study the evolution
of dwarf satellites directly in cosmological simulations rely on semi-analytical
methods to model the baryonic component rather than solving the fluid equations$^{18}$. These models 
neglect ram pressure, and since stripping by tides is slow and inefficient, they cannot explain the 
complete absence of gas and early truncation of star formation of the darkest dSphs.

In cold dark matter models the present-day spatial distribution of subhalos within primary halos 
retains some memory of their infall time$^{19}$. Satellites orbiting closer to the primaries 
were on average accreted earlier than those orbiting at larger distances. 
Interestingly, Draco and Ursa Minor lie at 68 and 86 kpc, respectively from the Milky Way, Andromeda IX at 45 kpc from
Andromeda, while other dSphs orbit as far as 200 kpc from the primaries$^1$.
We use a high resolution {\LCDM} dark matter-only 
cosmological simulation of the formation of a Milky Way-sized halo$^{20}$(see also Supplementary Information). 
At $z=0$ we identify three subhalos having distances below $100$ kpc from the center 
and with a peak circular velocity  in the range $25-30$ km/s$^6$
(Figure 1). We track the orbits of the satellites back in time and 
find that two of them were accreted early, between $z=2.5$ and $z=1.5$.

\begin{figure}
\vskip 10.8cm 
{\includegraphics{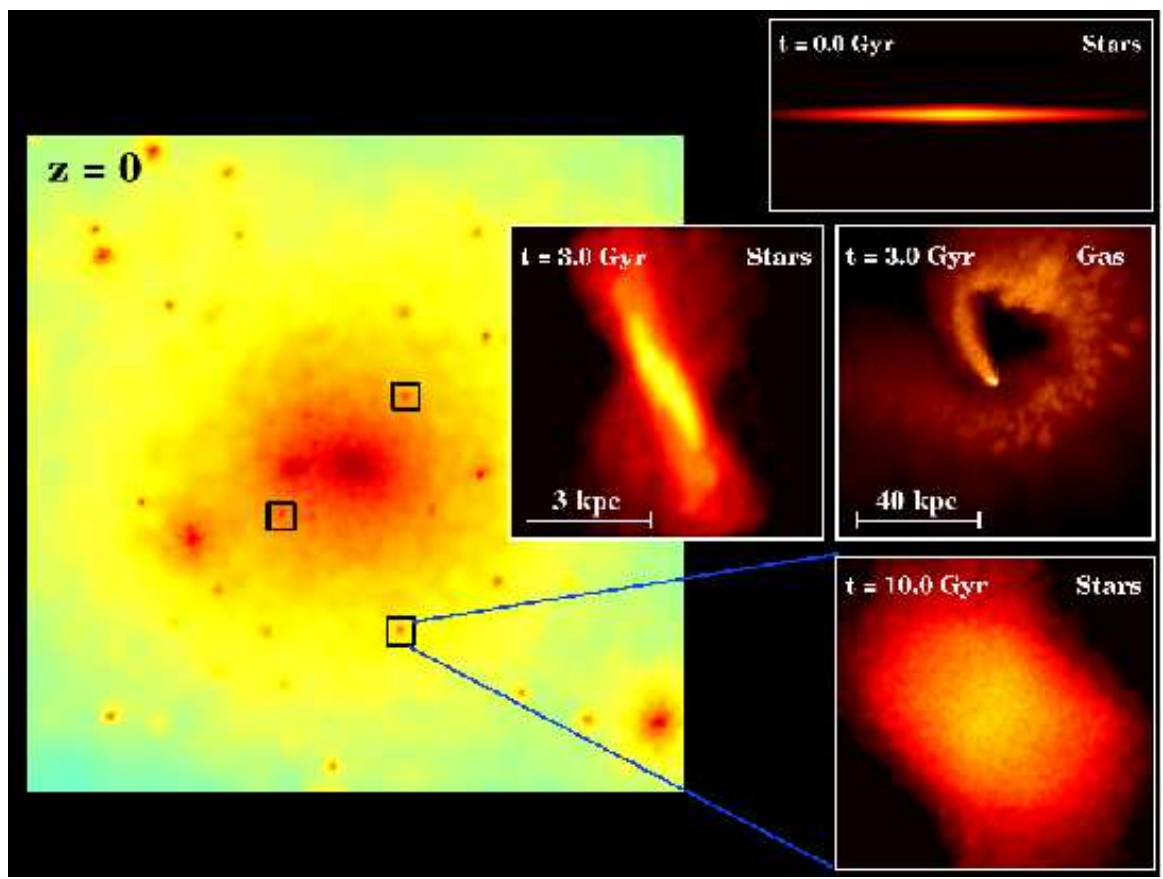}}
\caption[]{\small {\bf Morphological evolution of the dwarf satellite galaxy.}
{\it Left:} Color-coded projected density map of the cosmological run at $z=0$; the box is 260 kpc on a side, which corresponds to the 
virial radius of the Milky Way-sized halo. The peak density along the line
of sight is shown, ranging from $10^{-29} - 10^{-24}$ g/cm$^3$, with the color coding from blue 
(lowest density), through yellow, then red, to brown (highest density). 
The three satellites that meet the distance and circular velocity constraints (see text) are highlighted with black boxes. 
{\it Right:} From top to bottom the insets show the stellar component of the dwarf galaxy, color coded in projected density,
at different times. Only regions with densities in the range $10^{-28} - 10^{-23}$ g/cm$^3$
are shown, with the color coding from dark red (lowest density) to yellow (highest densities).
In the top inset, the initial disk is shown edge-on (the box is 8 kpc on a side). 
In the left middle inset, the system is
close to the second pericenter passage; the stars have assumed a strong bar-like
configuration and heating is evident in the outskirts (the box is 7 kpc on a side). In the bottom inset, the end state is shown;
the bar has been heated into a diffuse spheroid and any disk-like feature has been erased (the box is 4 kpc on a side and a
a projection along a random line of sight is shown). 
In the right middle inset, the trail of
gas produced by ram pressure is shown, while even the residual gas in the center is stripped. 
The color-coded gas density projected onto the orbital plane is shown (densities in the range $10^{-30} - 10^{-23}$ g/cm$^3$, color 
coding as above), and the box is 100 kpc on a side.}
\end{figure}

We then construct a high resolution N-body + smoothed particle hydrodynamics (SPH) model of a dwarf galaxy 
satellite having a disk of stars and gas inside a cold dark matter halo (Figure 1)
with a peak velocity of about $40$ km/s, comparable 
to that of the two identified cosmological subhalos before they 
were accreted onto the Milky Way (see Supplementary Information).
We assume that 80\% of the baryonic disk mass is in a gas component. The inefficient conversion of gas into
stars is expected at these low mass scales since most of the gas will have densities below the threshold
for star formation$^{21}$. In addition, at $z > 2$ the gas in the dwarf is heated to a temperature of over
$10^4$ K and ionized by the cosmic UV radiation, which further suppresses star formation.

The dwarf model is placed on an eccentric orbit inside a massive Milky Way-sized halo model which is a replica of that in the cosmological simulation. 
We include radiative cooling as well as the heating and ionizing flux from the cosmic ultraviolet background 
radiation$^{22}$, and we embed a diffuse gaseous halo inside the dark halo of the primary. Such a halo 
is expected as 
a by-product of the process of galaxy formation and has a density and temperature consistent with observational 
constraints (see Supplementary Information).

With an orbital time of about 1.7 Gyr the dwarf undergoes as many
as 5 pericenter passages in 10 Gyr. At the first pericenter passage its dark halo loses 60\% of its mass. The
disk, deep inside the potential well of the halo, suffers no stripping, but the tidal perturbation 
triggers a strong bar instability (Figure 1) and simultaneously heats the stars in the disk.
The bar funnels most of the gas towards the central kiloparsec.
Gas will be removed from the galaxy if the ram pressure force exerted by the diffuse hot gas,
which is proportional to $\rho_{\rm g} V^2$ ($\rho_{\rm g}$ being the density of the gaseous halo and $V$ being
the orbital speed of the galaxy), exceeds the gravitational restoring force provided by the
potential well of the galaxy$^{23}$.
Ram pressure readily removes the gas outside the bar radius on a timescale of less than $10^8$ years, but not 
the more tightly bound gas inside the bar.
When the satellite crosses the pericenter a second time, the new tidal shock lowers the
halo density by a factor of 2 inside 1 kiloparsec, so that $V_{\rm peak}$ drops to less 
than $30$ km/s.
Since the potential well has become shallower, even the gas sitting inside the bar can be swept 
away by ram pressure (Figure 1, see also Supplementary Information). The cosmic UV background 
heats and ionizes the gas, which enhances stripping significantly since the higher gas pressure opposes 
the gravitational restoring force. Once the first two orbits have been completed no gas is retained by 
the dwarf.

The response of the system to the tides becomes progressively more impulsive at each new pericenter
passage. As a result, the initially disky
stellar distribution is heated into a nearly spherical, isotropic configuration (Figure 1). The expansion 
reflects the attempt of the system to gain a new equilibrium as the internal binding energy is 
lowered. The removal of gas due to ram pressure is crucial for tidal heating to be effective 
(see Supplementary Information).

After 10 billion years a diffuse spheroidal galaxy has replaced the
gas-rich disk. During the last few orbits the stellar velocity dispersion profile is fairly flat (left panel of Figure 2), 
matching the profiles observed for Draco and Ursa Minor$^{24}$. The initial angular momentum of the
stars has been transported outward during the morphological transformation$^8$, producing a ratio between
rotational velocity and random motions $v_{\rm rot}/\sigma < 0.2$ in 
the inner few hundred parsecs.
The surface brightness 
and total luminosity resemble those of the faintest dSphs (see caption of Figure 2). 
A substantial halo is preserved within a few kiloparsecs
from the center so that the final $M/L$ is larger than 100 (right panel of Figure 2).
The central dark matter density, which has decreased by almost a factor of 4 since the
beginning of the simulation, is $\sim 0.2 M_{\odot}/$pc$^3$, comparable to that of
Draco and Ursa Minor$^1$.

\begin{figure}[t]
\vskip 8.0cm
{\includegraphics{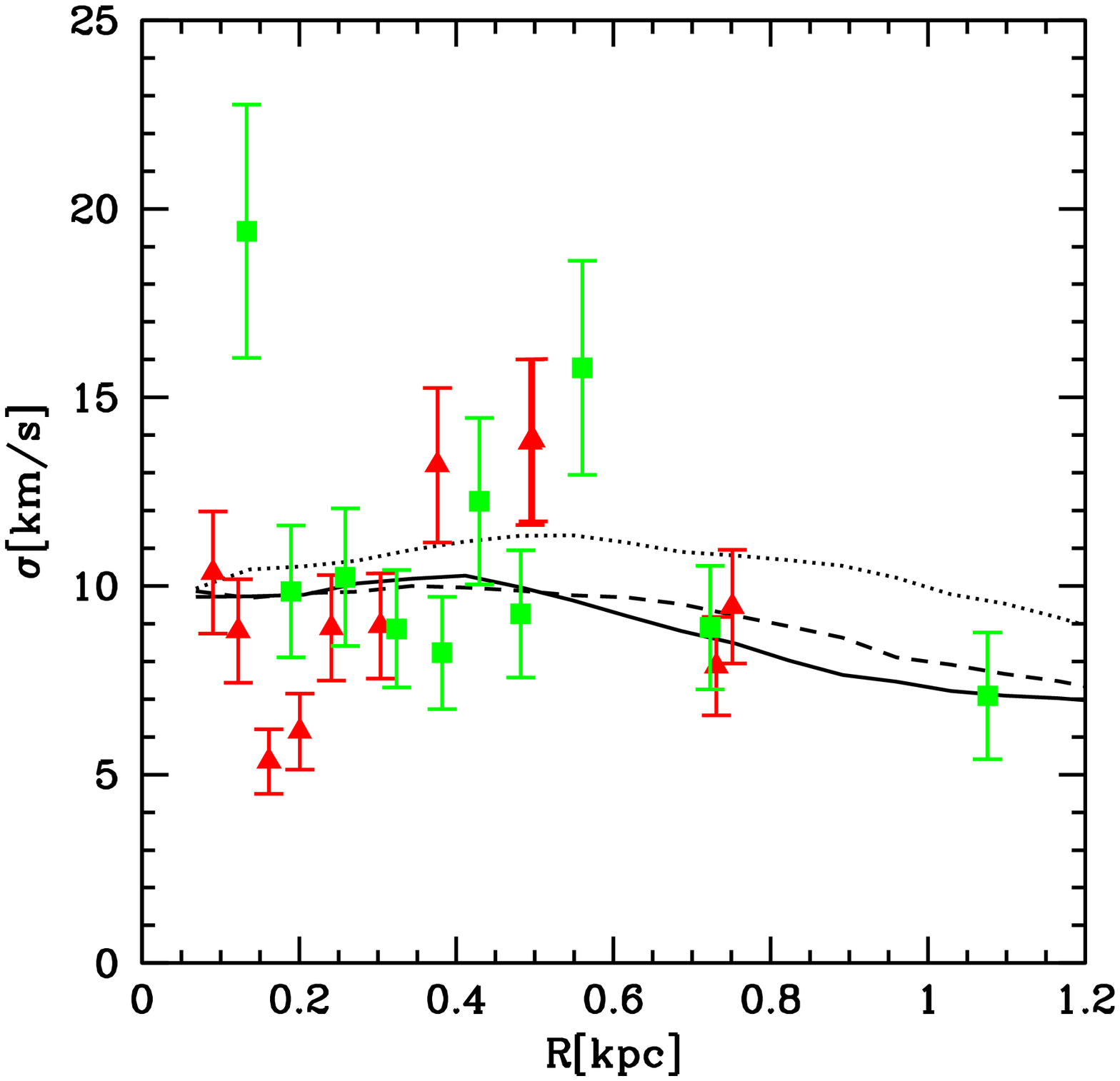}}
{\includegraphics{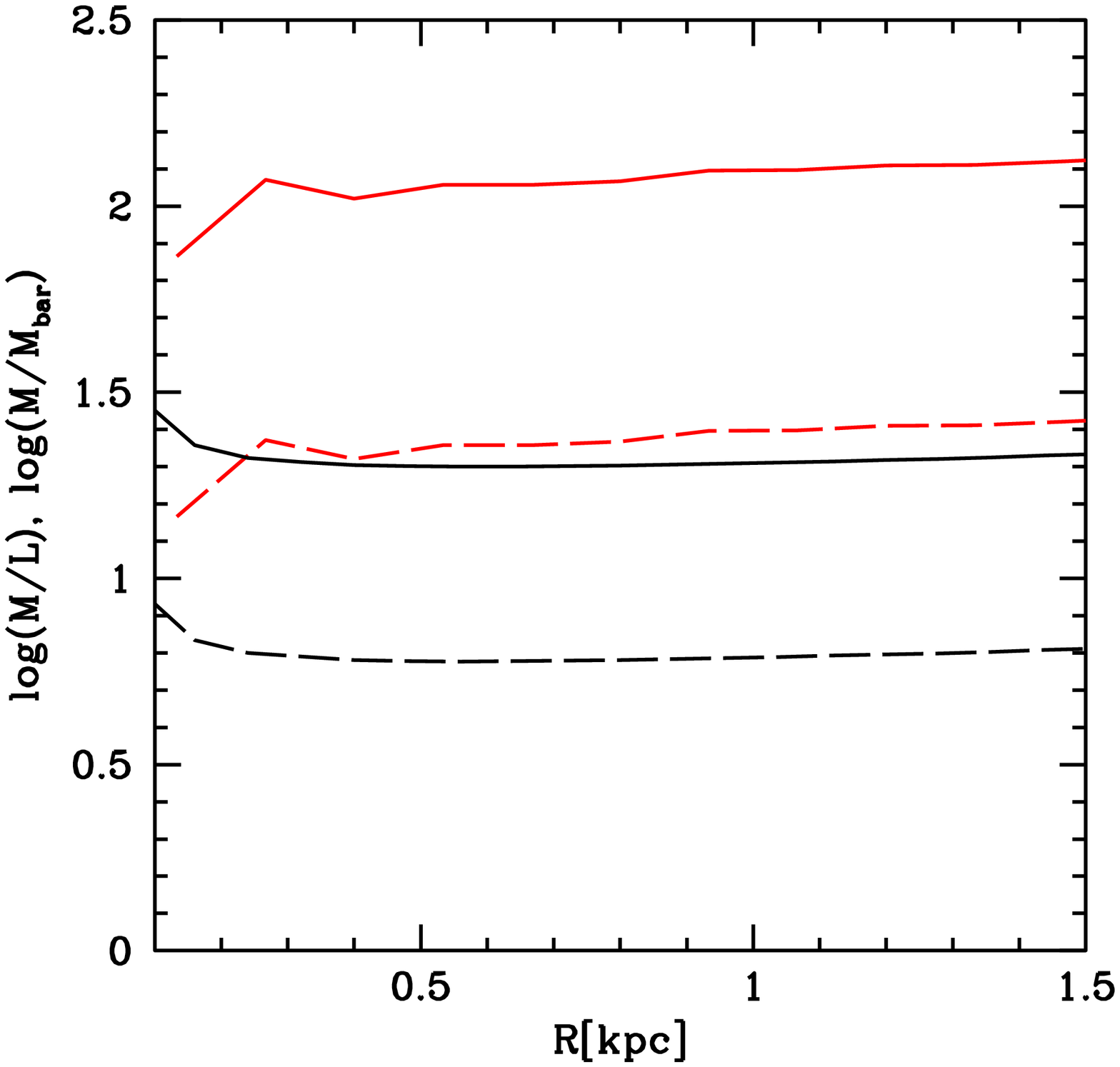}}
\caption[]{\small {\bf Structural properties of the simulated dwarf after 10 billion years of evolution.}
{\it Left:}  The line-of-sight stellar velocity dispersion profiles are shown 
for three random directions (black lines) perpendicular to each other, together 
with published observational data points for Draco (red triangles) and Ursa Minor (green squares), including
formal 1-$\sigma$ error bars $^{24}$.
The curves are shown out to the radius for which data points are available.
{\it Right:} Mass profiles shown out to the radius (from the center of the simulated dwarf) 
at which stars are gravitationally bound. 
The dashed lines show the initial (black dashed line) and final (red dashed line) ratios of the total
mass to the baryonic mass. The solid lines show the initial (black solid line) and final 
(red solid line) B band mass-to-light ratios of the dwarf.
We have assumed stellar mass-to-light ratio $(M/L_B)_*=1.5$ (initial) and  
$(M/L_B)_*=5$ (final) to compute the initial and final luminosities,
$M_B = -12.5$ and $M_B = -9$.  The final central surface brightness 
is $\mu_B \approx 26$ mag arcsec$^{-2}$. We note that an initial $(M/L_B)_*=1.5$ 
is motivated by the fact that at $z > 2$ a stellar population is at most 
three billion years old, whereas a final $(M/L_B)_*=5$ is consistent with
passive fading of the stellar population for about ten billion years$^{10}$.}
\end{figure}

We predict that all massive galaxies should 
have a few extremely dark matter dominated satellites as the mechanism 
reported here is completely general within hierarchical structure formation.
The efficiency of star formation in isolated low-mass galaxies can be lower than that assumed here$^{21}$ 
(see Supplementary Information), and hence dwarf satellites having gas fractions higher than
$80\%$ of the baryons before stripping probably did exist. Once they lose their gas, these systems will turn into dSphs
even more dark matter dominated than Draco and Ursa Minor. They should orbit
the Milky Way and M31 and have escaped detection so far owing to their low surface brightness, 
especially at low Galactic latitude$^{25}$. 
Their halo masses should be comparable to those of known dSphs and thus might help to solve the missing
satellites problem$^{26}$.  The Ursa Major dwarf recently detected by the Sloan Digital Sky survey
$^{27,28}$, which has a halo as massive as that of Draco but a luminosity a hundred times smaller, is probably
one of them. Three more systems such as Ursa Major
would be enough to bring theory and observations in agreement at the high end of the mass function
of satellites, corresponding to a peak circular velocity of around $25-30$ km/s, where suppression of baryonic infall by 
reionization is hardly effective.

When the ultraviolet background is an order of magnitude weaker, as is predicted$^{22}$ at $z <1$, about 30\% of the original gas component stays 
in the galaxy (see Supplementary Information).
Therefore dwarfs that fell into the Milky Way halo late can continue forming stars, and tidal shocks will produce periodic bursts of star formation$^{10}$. 
These newcomers should account for those dSphs that have fairly normal mass-to-light ratios, extended star 
formation histories and larger distances from the primaries$^{1,2}$.
This explains why Fornax is ten times brighter than Draco and has a very different star formation history despite 
having a comparable depth of the potential well$^{1,6,29}$.  
It implies that there should be a positive correlation 
between $M/L$ and the infall epoch of dwarfs, and thus a negative correlation between $M/L$ and their 
orbital time. The dSph Tucana represents the biggest challenge to our model because it lies far from any massive galaxy (see Supplementary Information). 
The accurate determination of the orbits of the dwarfs expected from on going and future astrometric missions such as the Space Inteferometry Mission 
and the Global Astrometric Interferometer for Astrophysics will be able to test this prediction.

\vskip 0.5cm

\begin{acknowledge}
Stimulating discussions with Y. Birinboim, J. Bullock, A. Dekel, F. Governato, A. Kravtsov, G. Lake, C. Porciani, J. Penarrubia, M. Valluri,
B. Willman, and A. Zentner are greatly acknowledged. We thank Steve Majewski and Ricardo Munoz for 
sharing their data with us. This research has been supported by the Zwicky Prize Fellowship program at the Swiss Federal Institute of Technology in 
Zurich and by the U.S. Department of Energy through a KIPAC Fellowship at Stanford University and the Stanford Linear Accelerator Center. This 
project was also supported in its initial stages by the Swiss National Science Foundation and the Kavli Institute for Cosmological Physics (KICP) 
at the University of Chicago. L. Mayer acknowledges KICP for hospitality during the initial stages of this project. 
L. Mayer and S. Kazantzidis are grateful to the Aspen Center for Physics for an invitation to the summer workshop ``Deconstructing the 
Local Group -- Dissecting Galaxy Formation in our Own Background'' where some of this work was completed. All computations 
were performed on the Zbox  supercomputer at the University of Z\"urich, on LeMieux at the Pittsburgh Supercomputing Center, and 
on the Gonzales cluster at ETH  Z\"urich.

\end{acknowledge}

\vskip 0.5cm
\noindent{\bf Author Information} The authors declare that they have no competing financial
interests. Correspondence and requests for materials should be addressed to L.M. 
(lucio@phys.ethz.ch) or S.K. (stelios@slac.stanford.edu).
\newpage


\begin{center}
\vspace*{-10.00pt}
{\bf SUPPLEMENTARY INFORMATION}
\end{center}

\noindent Here we briefly describe the setup of the initial conditions and the 
numerical methods used to perform the simulations presented
in the Letter. This is followed by a discussion of the tests and the analysis performed in
order to verify and understand the results in depth.

\section {Simulation code}

We have used the fully parallel, N-body+smoothed particle hydrodynamics (SPH) 
code GASOLINE to compute the evolution of both the collisionless and 
dissipative component in the simulations. A detailed description of the code is available in the literature$^1$. Here we recall
its essential features. 
GASOLINE computes gravitational forces 
using a tree--code$^2$ that employs multipole expansions to approximate the gravitational
acceleration on each particle. A tree is built with each node storing
its multipole moments.  Each node is recursively divided into smaller
subvolumes until the final leaf nodes are reached.  Starting from the
root node and moving level by level toward the leaves of the tree, we
obtain a progressively more detailed representation of the underlying
mass distribution. In calculating the force on a particle, we can
tolerate a cruder representation of the more distant particles leading
to an $O(N \log{N})$ method. Since we only need a crude representation 
for distant mass, the concept of ``computational locality'' translates 
directly to spatial locality and leads to a natural domain decomposition.  
Time integration is carried out using the leapfrog method, which is a 
second-order symplectic integrator requiring only one costly force 
evaluation per timestep and only one copy of the physical state of the system.
In cosmological simulations  periodic boundary conditions are
mandatory. GASOLINE uses a  generalized  Ewald method$^3$ to arbitrary
order, implemented through hexadecapole. 

SPH is a technique of using particles to integrate 
fluid elements representing gas$^{4,5}$.
GASOLINE is fully Lagrangian, spatially and temporally adaptive and efficient
for large $N$. The version of the code used in this Letter includes 
radiative cooling and accounts for the effect of a 
uniform background radiation field 
on the ionization and excitation state of the gas. The cosmic ultraviolet 
background is implemented using the Haardt-Madau model$^7$ which
includes photoionizing and photoheating rates produced by QSOs and galaxies
starting at $z=7$. The assumption that reionization occurred at $z=7$ is conservative 
but consistent with the combination of the 3rd year  WMAP results and the Gunn-Peterson effect in the
spectra of distant quasars$^8$. We use a standard cooling 
function for a primordial mixture of atomic hydrogen and helium 
(the metallicity in dSphs is indeed much lower than solar$^9$, with 
$-1 <  [Fe/H]  < -2$). The
cooling shuts off below $10^4$ K. The internal energy of the gas is 
integrated using the asymmetric formulation, that gives results very close 
to the entropy conserving formulation$^6$ but 
conserves energy better. Dissipation in shocks is modeled using the 
quadratic term of the standard Monaghan artificial viscosity$^5$. 
The Balsara correction term is used to reduce unwanted shear viscosity$^{10}$.

\section {Initial Conditions: the cosmological simulation}

The cosmological simulation employs the so called volume renormalization
technique to achieve higher resolution in a region of interest (described 
in the next section). It is one of the 
simulations already presented elsewhere$^{11, 16}$. The virial mass of the halo that was selected at $z=0$ to be 
re-simulated at higher resolution is $1.15 \times 10^{12} M_{\odot}$ (the
virial mass is  measured within the virial radius R$_{\rm vir}$, the
radius enclosing an overdensity of 100 times the critical density
$\rho_{\rm crit}$). A recent analysis of all observational constraints 
points to a Milky Way halo of about $10^{12} M_{\odot}$$^{11}$.
The halo was originally selected  within a low resolution, 
dark matter only simulation run in a concordance, flat, $\Lambda$-dominated
cosmology: $\Omega_0=0.3$, $\Lambda$=0.7, $h=0.7$, $\sigma_8=0.9$,
shape parameter $\Gamma=0.21$, and $\Omega_{b}=0.039$$^{12}$.  The size of the box, 100 Mpc, 
is large enough to provide realistic torques.  The power spectra to model the
initial linear density field were calculated using the CMBFAST code
to generate transfer functions.
Dark matter particle masses in the high resolution regions are 8.05 $\times$
10$^5$ $M_{\odot}$, and and the force resolution, i.e. the gravitational softening, is
0.3 kpc. In total there are $1.4 \times 10^6$ particles within the virial
radius.
 
With our choices  of particle number and softening, the smallest subhalos resolved have typical circular velocities of 10\% of their host.  For all particle
species in the high resolution region, the gravitational spline
softening, $\epsilon(z)$, evolved in a comoving manner from the starting
redshift (z $\sim$ 100) until z=9, and remained fixed at its final
value from z=9 to the present. The softening values chosen are a good
compromise between reducing two body relaxation and ensuring that
disk scale lengths and the central part of dark matter halos will be
spatially resolved. Integration parameter values were chosen according
to the results of previous systematic parameter studies$^{14}$.

The merging histories and angular momentum of the parent dark matter halo
are supposed to play a major role in defining the final properties of the galaxies that form inside them. It is then important to make
sure that our halos have merging histories and spin parameters
somewhat representative of the global population.  The halo was selected
with the only criteria that the redshift of the last
major merger (z$_{lmm}$) was $>$ 2 and that there are no 
halos of similar or larger
mass within a few virial radii (a major merger is defined here as
having a 3:1 mass ratio). The halo spin parameter, $\lambda$,
is $\sim 0.05$ at $z=0$, close to the  average value for cosmic halos
$\sim$ 0.035$^{15}$. 
Cosmological simulations with hydrodynamics
that use this same and other similar initial conditions show that realistic 
disk galaxies similar to the Milky Way do arise in these dark matter halos$^{16}$.

\subsection{Volume renormalization technique}

The large dynamic range involved in cosmological simulations aimed at
resolving the scale of galactic halos calls for techniques that concentrate
the computational power on the object of interest. This is achieved by
the volume renormalization technique$^{17}$. 
A large scale simulation is done at
low resolution and the regions of interest are identified, e.g. QSOs
forming at high redshift, a large cluster of galaxies or a Milky Way-sized
halo such as in this Letter. Next, initial conditions are 
reconstructed using the same low-frequency waves present in the low resolution
simulation but adding the higher spatial frequencies.  To reduce the
number of particles and make the nonlinear simulation possible with
the same cosmological context, we construct another set of initial
conditions with particles whose mass, and therefore mean separation,
increase with the distance from the center of our volume.  Note that
because tides are important in the formation of the filaments it is
not sufficient to extract just the central region.
Generating initial conditions (ICs) in this approach is nontrivial.
The ICs are calculated on a grid determined by the initial mass
resolution.  The power spectrum is realized by using Fast Fourier Transforms
(FFTs) to determine
displacements of particles from this grid.  We use a 
particular technique that allows one to calculate high resolution FFTs 
only in the regions of the  simulations were mass and force resolution 
need to be high. This method has been successfully used in a wide
range of cosmological studies$^{18,19, 20, 11}$.

\section {Initial Conditions: the dwarf galaxy model}

The multi-component dwarf galaxy models are built using a standard technique
that produces initial conditions very close to equilibrium$^{20, 21}$.
The model is further evolved in isolation for several Gyr in order to remove
possible transients caused by initial particle noise, and the end state is used as initial condition. 
The scheme we use to assign the structural parameters of the 
baryons in a dark matter halo assumes that baryons collapse into the halo
having the same specific angular momentum distribution of the halo and 
conserving angular momentum during collapse$^{23}$.
The properties of the baryonic
component are related to those of the dark halo, namely its mass, angular 
momentum (via the spin parameter) and concentration. 
We refer to previous papers for a detailed 
description of the scaling relations that relate the baryonic component and 
the dark halo$^{24, 25}$. 
The technique has been widely used to build models of dwarf
galaxies$^{26, 27}$.

\begin{figure}[t]
\vskip 8cm
{\includegraphics{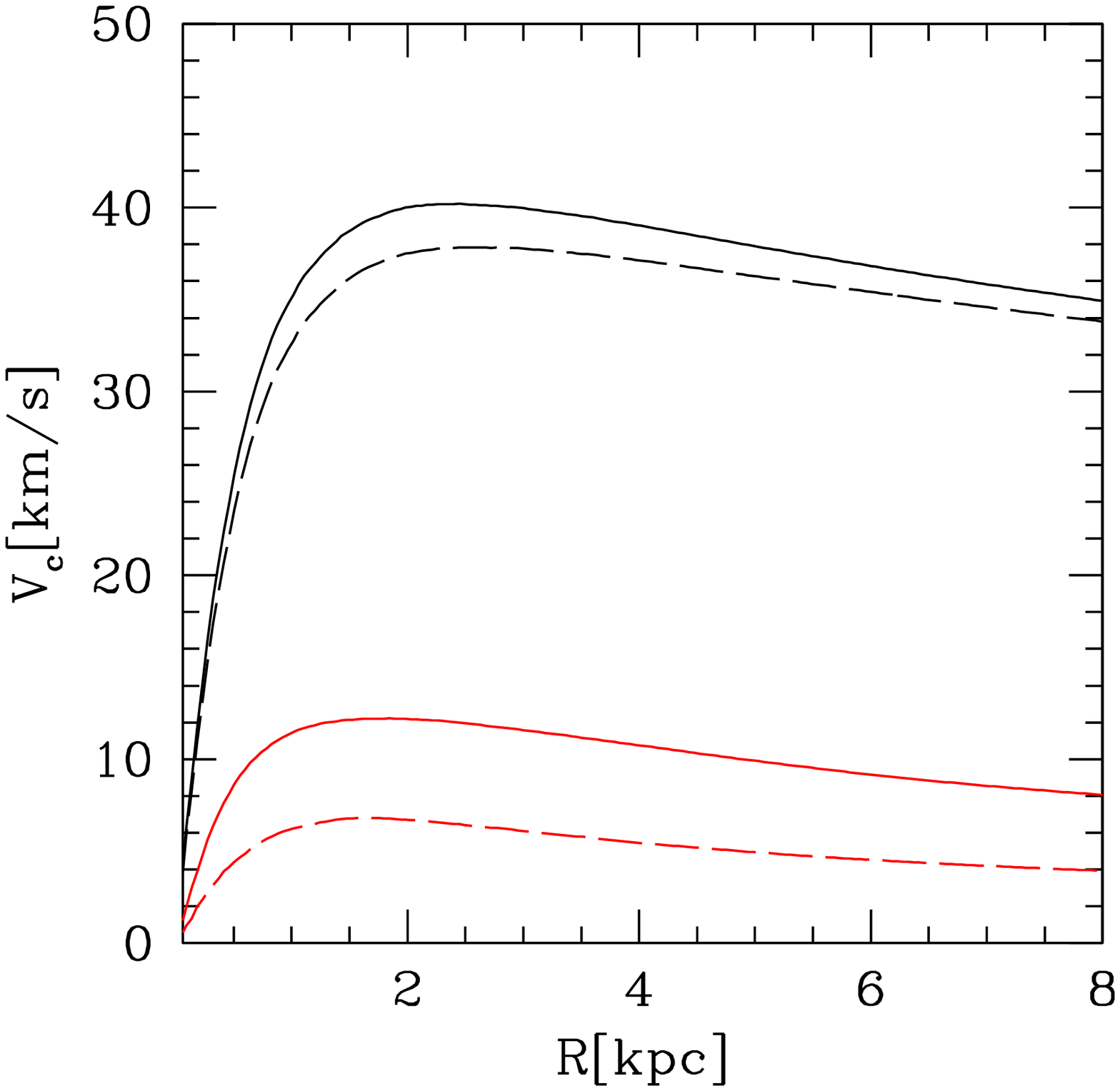}}
{\includegraphics{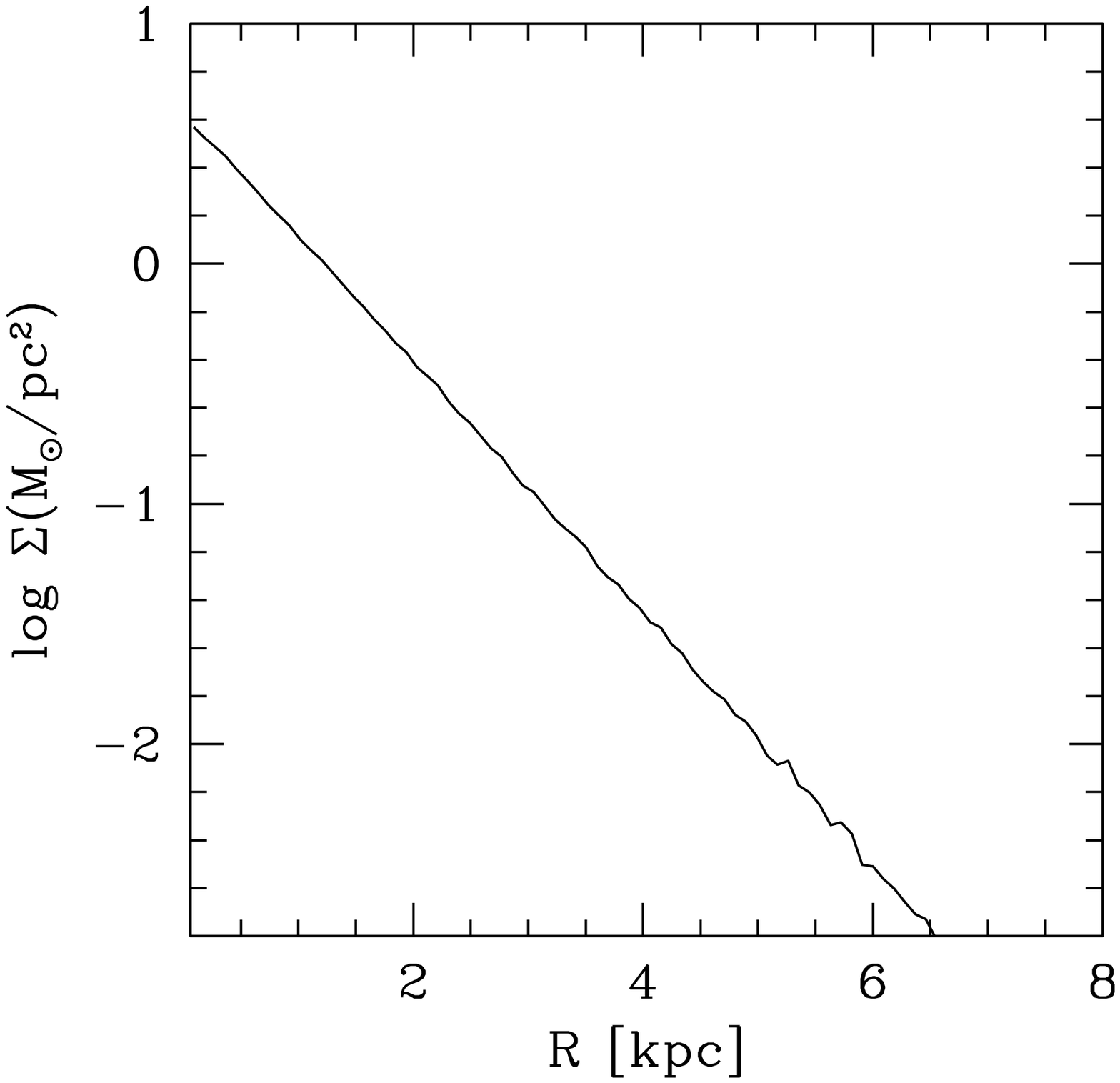}}
\caption[]{\small Structural properties of the initial dwarf galaxy model.
The left panel shows the initial rotation curve. The red solid line shows the
contribution of the gas, the red dashed line the contribution of the stars, the black dashed
line shows the dark matter contribution and the black solid line shows the total
rotation curve resulting from the sum of the contribution of the different components.
The right panel shows the initial baryonic surface density profile.
}
\end{figure}

The bayonic component comprises an exponential disk of gas and stars 
embedded in an NFW halo$^{28}$. The halo has $V_{\rm peak}=40$ km/s.
In CDM models low mass halos have on average higher concentration parameters
$c=R_{\rm vir}/r_{\rm s}$ (where $R_{\rm vir}$  is the virial radius and $r_{\rm s}$ the scale
radius) relative to higher mass halos and, for a given halo mass, the concentration
decreases with redshift roughly as $(1+z)^{-1}$. We choose
$c=20$ for the halo of the dwarf. This value of the halo concentration matches the average
concentration of the subhalos  with $V_{\rm peak}$ in the range $30-40$ km/s in the
cosmological simulation before they fall into the primary halo and is consistent with 
the typical values expected at these mass scales at $z \sim 2$, as shown elsewhere$^{29}$. 
In choosing the radius and mass of the disk there
is some freedom. However, there are constraints coming from the link between
disk radius and the spin parameter of the halo in which the disk forms,
and from the information of the typical fraction of the mass of present-day 
dwarf galaxies that is contributed by the baryonic disk.
The first 
constraint brackets the range of possible disk radii at given disk mass, 
the second the range of possible disk masses for a given halo mass. 
Since our simple working hypothesis is that the progenitors of dwarf
spheroidals had baryon 
fractions in the range of the normal values of galaxies  
in the present-day Universe, we fix the disk mass fraction to 
$f_d = 0.035 M_{\rm vir}$.
This value is typical in mass models that reproduce dwarf and low surface 
brightness
galaxies$^{29}$ and is consistent with the baryonic Tully-Fisher
relation for galaxies with rotational speeds in the range $40-60$ km/s$^{30}$.
The exponential disk scale length is then fixed by the adopted spin parameter,
$\lambda=0.05$. The log-normal distribution of spin parameters of 
cosmological halos peaks at around $0.035$$^{15}$ but (1) 40\% of the 
halos have larger spins and (2) modeling of their rotation curves suggests 
that dwarf galaxies have an average spin larger than the mean value of the  
galaxy population as a whole$^{29}$.
The resulting exponential disk scale length, which is a function of
$\lambda$, the halo concentration $c$ and $f_{\rm d}$$^{24}$,
is $0.55$ kpc.

The chosen disk mass and radius yield an initial central baryonic surface 
density $\sim 4 M_{\odot}/$pc$^2$. 
We choose the stellar mass fraction in such a way that it is consistent
with the amount of star formation expected in a galaxy with the assigned
structural parameters. Within the standard structure formation paradigm galaxies in halos
of $10^{10} M_{\odot}$ or less, namely in the range of our model, should
have little or no star formation because their disks are gravitationally stable
nearly everywhere$^{31}$, having Toomre parameters $Q > 1$. The stability arises because lower 
mass halos should harbor lighter disks. The latter statement is true if the disk mass fraction 
either varies little or tends to decrease for decreasing halo masses, in agreement with the analysis 
of galaxy rotation curves across a range of masses$^{30,31}$.
Observations show examples of galaxies that are Toomre-stable and yet have on-going
or recent star formation, albeit at a fairly low level$^{32,33}$.
Recent models can account even for such departures from the Toomre
criterion$^{34,35}$. In one such model the threshold for star formation
is determined  by the minimum surface density required for molecular gas to form$^{34}$.
In our dwarf galaxy model about  $1.5 \times 10^7 M_{\odot}$ of baryons have surface densities above
the predicted critical surface density predicted, which is  $\sim 2  M_{\odot}/$pc$^2$. 
We thus assign to our model a stellar mass of 
$1.5 \times 10^7 M_{\odot}$, corresponding to only 20\% of the baryons.
Since the critical surface density 
for star formation can rise by up to a factor of 5 in presence of 
ionizing radiation fields such as the cosmic ultraviolet background$^{34}$, 
the assumed stellar mass fraction should be viewed as an upper limit for 
a reasonable progenitor at these mass scales. 
Although the criterion based on surface density is convenient,
it is known to have difficulties in reproducing some
observations. Therefore we compared our choice with another, more recent work that
predicts the stellar fraction using fully self-consistent N-body+SPH simulations with
a physically motivated star formation recipe and sink particles to represent sites where
molecular clouds and star clusters would form$^{35}$. These simulations employ models
of galactic disks with a variety of initial structural parameters, gas mass fractions
and gas temperatures. They explore the dependence of the star formation rate on
the initial minimum Toomre parameter of the combined gaseous and stellar disk, $Q_{sg}$,
and find that star formation is possible up to values of $Q_{sg} \sim 1.6$, significantly
larger than the standard threshold $Q_g \sim 1$, where $Q_g$ is the local Toomre parameter of 
the gas. Moreover, by running a large set of simulations for about 2 Gyr, they derive a relation 
between the minimum initial Toomre parameters of the systems, $Q_{sg}(min)$ or $Q_g(min)$, and the 
total amount of gas that can be converted into stars. We compared our initial setup with their results.
Assuming a fully gaseous disk at the start, our initial galaxy model would have 
$Q_{sg}(min) = Q_g(min) \sim 1.42$ (reached between 1 and 2 disk 
scale lengths). For such a value of $Q_g(min)$, Figure 15 of their paper (bottom panel)
indicates that between 10\% and 17\% of the initial gas should be converted into stars
(13\% if we compare with their best fitting curve). 
These fractions are close to the 20\% that we assumed, though somewhat smaller; indeed, as we emphasize 
in the Letter, a smaller initial stellar fraction would make our scenario even stronger.
The rotation curve and initial baryonic surface density profile
of the dwarf galaxy model are shown in Figure 3.

We choose the mass and force resolution in the dwarf galaxy models in such a way that possible
numerical artifacts are under control.
We adopt $2\times 10^5$ particles in the gaseous disk and $10^6$ particles in the stellar disk
and dark matter halo, for a total of $2.2 \times 10^6$ particles.
The spatial resolution of the dark halo and stars is set by the gravitational softening which
is equal to 35 pc.
The spatial resolution in the gas component is given by the SPH smoothing length, defined as
the radius of the spherical kernel adopted by GASOLINE, which contains 32 particles. The smoothing
length is adaptive in space and time. In the initial conditions it is about 25 pc 
at about one disk scale length, and it drops to $\sim 15$ pc during the stage of the bar driven gas inflow 
close to the first pericenter passage.
Since most of the mass in an exponential disk is contained within about 3 scale lengths, or 
1.5 kpc, the gaseous disk is resolved by about 60 resolution elements initially; 
later, as the bar forms, the average smoothing length decreases, but also the size of the gas component 
becomes smaller by almost a factor of 2 (due to the gas inflow and stripping of the outer regions), so 
that effectively the resolution remains close to 60 smoothing kernels. 
This resolution is above that required to properly model ram pressure stripping  based on previous 
works which found that 50 resolution elements in a linear dimension (or 25 along a radius) 
is needed$^{36,37}$. 
Moreover, both the smoothing length and the gravitational softening of the gas are at all times
much smaller than the local Jeans length, and have comparable values.
This ensures that gravity and pressure forces are 
correctly balanced at the  Jeans length, thus avoiding spurious fragmentation or artificial suppression
of collapse$^{38}$. The Jeans length is defined as $\lambda_J = \sqrt{\pi v_s^2 \over G \rho_0}$, 
where $v_s$ is the sound speed, $G$ is the gravitational constant and $\rho_0$  is the (local)
density. It is smallest during bar formation
since the temperature of the bound gas is $\sim 2 \times 10^4$ K at all times while the density peaks
at this time, being close to $10^{-23}$ g/cm$^3$ near the center. Yet even in this extreme situation
$\lambda_J \sim 300$ pc, i.e. about an order of magnitude larger than either twice the SPH smoothing
length or the gravitational softening$^{38}$.

\section{Initial Conditions: the Milky Way halo model}

The live dark halo model used for the Milky Way is based on the Milky-Way 
sized halo of the cosmological simulation (see above). We use a 
slightly larger value of the circular velocity at the
virial radius, $V_{\rm vir}=180$ km/s, in order to have 
$V_{\rm peak}=230$ km/s at about 10 kpc from the center without
including a baryonic disk component. The concentration is $c=10$.
With our choice of parameters the shape of the
rotation curve is quite flat out to $100$ kpc and the resulting halo mass 
within $100$ kpc is 
$\sim 8 \times 10^{11} M_{\odot}$, consistent with 
the orbital dynamics of the Magellanic Clouds and distant dSphs$^{39, 40}$. 
We use $10^6$ particles to
model the halo and a softening of 300 pc, as in the cosmological simulation.
We embed a gaseous halo in hydrostatic equilibrium within the live dark 
matter halo. The density of the gaseous halo near the pericenter of the
orbit of the satellite, $30$ kpc, is $2 \times 10^{-4}$
cm$^{-3}$, comparable to that of the diffuse gas in recent hydrodynamical 
cosmological simulations that follow the formation of a disk galaxy inside 
the same halo used in this paper$^{16}$.
Indirect estimates of the present-day density of the gaseous corona 
surrounding the Milky Way based on OVI,OVII and X-ray absorption measurements
are also in the same range ($\sim 10^{-4}$ cm$^{-3}$ at a distance of 50 kpc$^{42}$). The temperature of the halo is $10^6$ K at 50
kpc, which is consistent with the same observational constraints.
Assuming an isotropic model, the halo temperature
at a given radius $r$ is determined by the cumulative mass distribution $M(r)$ of
the dark, stellar and gaseous components of the Milky Way beyond $r$ and by the density
profile $\rho_h(r)$ of the hot gas$^{27,43}$
\begin{equation}
T(r) = \frac{m_p}{k_B} \frac{1}{\rho_h(r)} \int_{r}^{\infty} \rho_h(r)\frac{GM(r)}{r^2} \, dr \, , 
\end{equation} where $m_p$ is the proton mass, $G$
and $k_B$ are the gravitational and Boltzmann constants.
In reality, the gaseous halo will not be in hydrostatic equilibrium since it would slowly 
cool and sink to the center while at the same time new gas flows within the 
virialized region from outside. However, cosmological hydrodynamical 
simulations$^{16}$ indicate that the characteristic density of the halo at $30-50$ kpc from the center does not change significantly from $z=2$ to $z=0$.

\begin{figure}
\vskip 13.2cm 
{\includegraphics{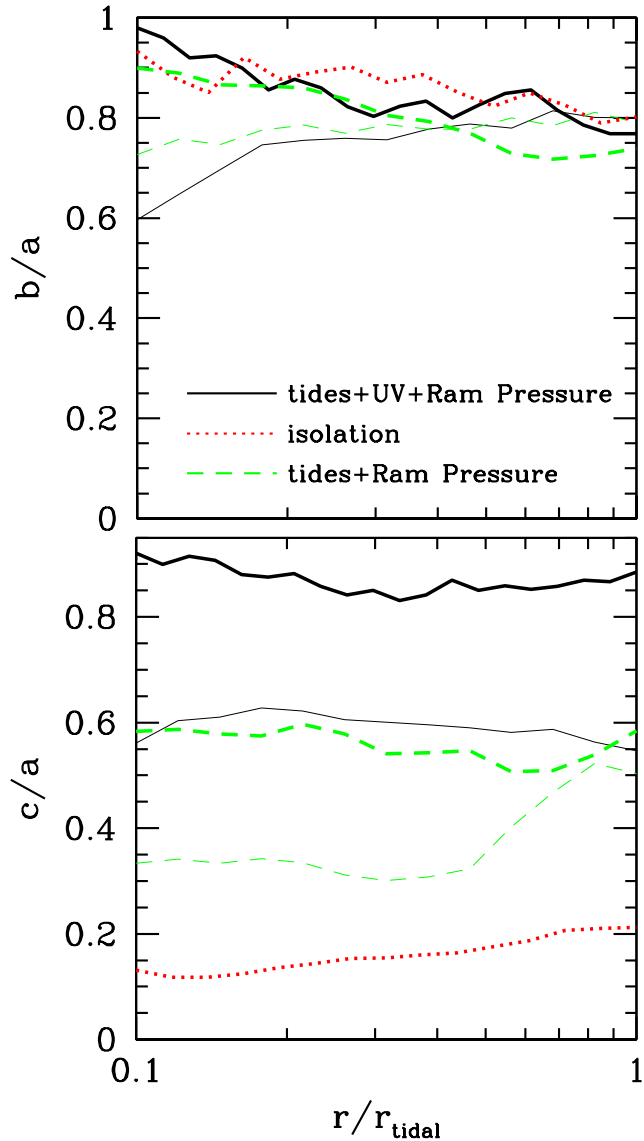}}
\caption[]{\small 
Shape evolution of the stellar component. The upper panel shows 
the intermediate-to-major axis ratio, $b/a$, and the bottom panel presents 
the minor-to-major axis ratio, $c/a$. Solid lines correspond to the self-consistent
simulation in which tides, ram pressure stripping, and UV background were included
throughout the orbital evolution of the dwarf. Dotted lines correspond to the case
where the dwarf galaxy was extracted from the simulation just after the 
gas was removed and evolved in isolation.
Dashed lines show results of a test in which approximately 30\% of the
initial gas mass remains in the galaxy because the absence of
the UV background reduces significantly the effectiveness of ram pressure
stripping. Thick and thin lines correspond to 10 and 5 Gyr of orbital
evolution, respectively (for the extracted dwarf we only show the result after
10 Gyr). The shape of the stellar component changes gradually over time
and the presence of UV heating is essential for the shape transformation of the 
dwarf.}
\end{figure}

\section{Tests of the results}

\subsection{Tidal heating and gas removal}

We have performed a number of tests to explore in detail the nature of the
transformation of the baryonic component. The main goal was to determine
the main cause of the expansion and roundening of the stellar component.
We investigated the relative role of the rapid removal of the gas, which comprises most
of the baryons initially, and that of the repeated tidal shocks in changing
the shape of the stellar component. Although the initial dwarf model is dark
matter dominated by construction, baryons comprise as much as 30\% of the total
mass within the initial exponential disk scale length. Their rapid removal
due to ram pressure stripping might in principle affect the dark matter potential,
possibly lowering its density as found in previous studies where baryonic 
stripping occurs as a result of powerful supernovae winds$^{43,44,45}$.

Figure 4 shows principle axis ratios $s = b/a$ and $q =
c/a$ ($a>b>c$) of stars, calculated from the eigenvalues of a modified inertia tensor$^{46}$
: $I_{ij} = \sum_{\alpha}
x_{i}^{\alpha}x_{j}^{\alpha}/r^{2}_{\alpha}$, where $x_{i}^{\alpha}$
is the $i$ coordinate of the $\alpha$th particle, $r^2_{\alpha} =
(y^{\alpha}_{1})^2 + (y^{\alpha}_{2}/s)^2 + (y^{\alpha}_{3}/q)^2$, and
$y^{\alpha}_{i}$ are coordinates with respect to the principle axes.
We use an iterative algorithm starting with a spherical
configuration ($a=b=c$) and use the results of the previous
iteration to define the principle axes of the next iteration until the
results converge to a fractional difference of $10^{-2}$.
Results are shown for three experiments. 
The comparison between the red and the black lines clearly shows that repeated tidal shocks are necessary for 
the transformation of the stellar component to occur;
the increase of $c/a$ is amplified by more than a factor of 6 when the dwarf
is continuously tidally shocked relative to a case in which it just loses 
most of its baryons owing to ram pressure and then evolves without any
tidal perturbation (in the latter case we removed the primary halo from the
simulation after the gas was stripped).
The green lines in Figure 4 show a test in which about 30\% of the initial gas
content is retained until the end 
because the UV background is absent and thus the effect of ram pressure is reduced
(we tested that not including the cosmic UV background at all or including it with the low amplitude expected at
$z < 1$ from the Haardt \& Madau model yields the same result$^{26}$).
The residual gas is funnelled to the center by the bar, where it forms a very dense
knot$^{26}$. We notice that despite the fact that this is a small fraction of 
the initial
gas content it represents a non-negligible contribution to the central potential because
it all ends up in the inner 0.5 kpc (corresponding to the initial
disk scale length). In this case the dwarf maintains a bar-like, prolate shape
even after 10 Gyr, hence the lower $c/a$.
After 5 Gyr the central density is 8 times higher compared to the simulation
in which complete gas stripping occurs. As a result $t_{\rm shock} > t_{\rm orb}$ within
0.5 kpc, where $t_{\rm shock}$ is $R_{\rm peri}/V_{\rm peri}$, $R_{\rm peri}$ and $V_{\rm peri}$ being, respectively,
the pericentric distance and the velocity at pericenter, and $t_{\rm orb}$ is the orbital time at
0.5 kpc from the center of the dwarf; 
the response of the system to the tidal forcing is thus adiabatic instead of 
impulsive, hence further morphological changes are inhibited.
This demonstrates that while gas removal does not drive the transformation it represents
a crucial step for the evolution of the 
stellar component because the effect of the tidal shocks is considerably weakened when a dense
central gas component is retained.

Figure 4 also shows that $c/a$ does not grow beyond $0.2$ in isolation, even after a few Gyr of evolution.
The average (apparent) ellipticity measured for late-type dwarfs 
in the Local Group is larger, $\sim 0.58$, for the 8 dwarfs having a range of rotational velocities comparable with
those of our initial dwarf galaxy model ($V_{\rm peak} \sim 30-40$ km/s)$^{47}$. Similar values are found for dwarf irregular
galaxies in clusters$^{48}$. However, in comparing with observed values of ellipticity one must take
into account inclination effects, since the $c/a$ discussed for our simulated dwarf is the intrinsic value, i.e.
the value that would be measured by an observer viewing the dwarf exactly edge-on. By viewing the dwarf at 
inclinations in the range 30-60 degrees, which statistically are a much more representative situation, the mean apparent 
$c/a$ varies in the range $\sim 0.4-0.7$ within 3 disk scale lengths (it is $0.55$ at 45 degrees of inclination), hence 
perfectly consistent with the observed values.

The central dark matter density decreases by almost a factor of 4 during the evolution as a result of the repeated 
tidal shocks when all the gas is readily removed.
Since the dwarf is dark matter dominated, it is natural to ask whether 
the change in the halo potential associated to this drop of the halo central density  
is sufficient to alter the mass distribution of the stellar component or if the direct
heating of the disk by the tidal shocks is crucial.
Figure 5 shows thickness of the stellar disk defined as the dispersion of particle
positions above the  midplane ($z=0$), as a function of cylindrical radius $R$.
The black line corresponds to the initial disk and red line corresponds to the disk
thickness after $5$ Gyr of evolution in isolation. The good agreement between 
these two curves suggests that the initial disk model is in reasonable equilibrium.
The green line shows instead the results of an experiment with the isolated dwarf galaxy model in which
 we decreased the mass of the inner dark matter halo over $5$ Gyr by the same amount as in the simulation in which the primary halo is included. 
The decrease of the central dark matter mass is of order of $40\%$ over $5$ Gyr and is responsible 
for an increase in thickness of less than $30\%$. 
The stellar component maintains an evident disk-like structure. Hence the transformation of the disk 
is not driven by the response to the changing dark matter potential.

In summary, these experiments indicate that the repeated tidal shocks in the impulsive regime 
are driving the transformation of the shape and mass distribution of the dwarf galaxy
and that gas removal is a necessary condition for them to be effective. 
We note that in previous works on tidal stirring$^{49,50}$ morphological changes were happening primarily 
as a result of tidally induced dynamical instabilities rather than impulsive heating. There the buckling of the bar was driving the change of shape from a disk 
into a spheroid. Here buckling never occurs because
the self-gravity of the stars is too low for global  bending modes to grow. In fact 
in those previous works the dwarf models had similar baryonic masses but stellar masses
nearly an order of magnitude larger (conversely the gas was at most 
contributing 30\% of the baryonic mass).

\begin{figure}[t]
\vskip 12.0cm 
{\includegraphics{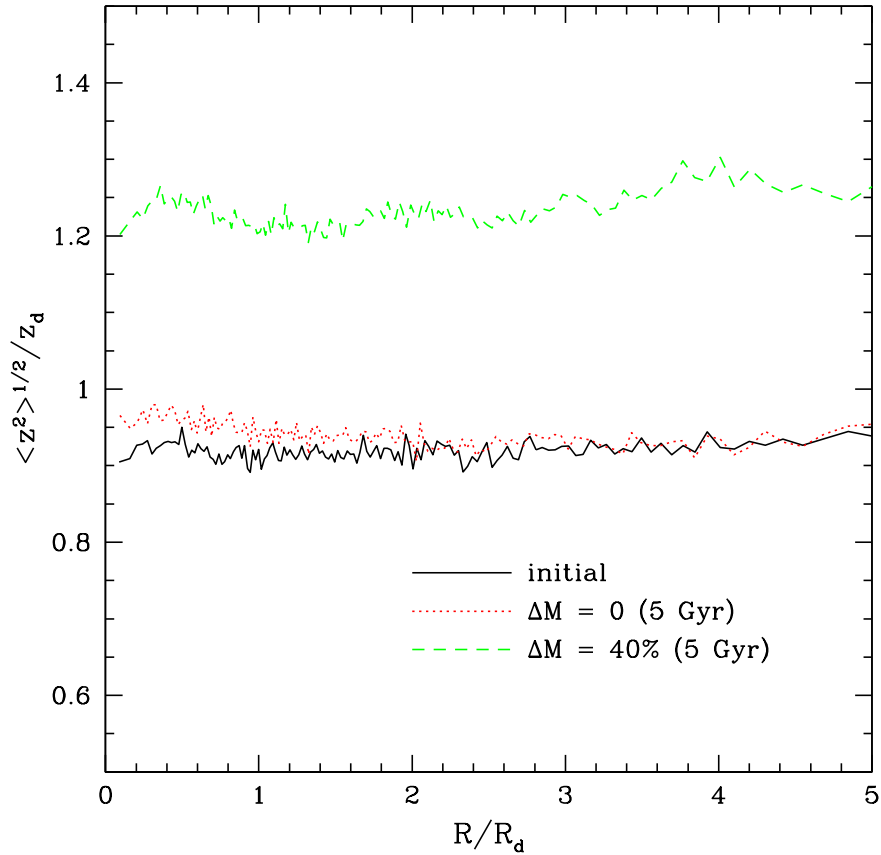}}
\caption[]{\small Evolution of the disk thickness as a function of radius, computed relative to the initial disk 
scale height, $z_d$ that characterizes the vertical surface density profile ($R_d$ is the initial disk scale
length). The results are shown for the run in isolation (red line) and for the run
in which 40\% of the halo mass within a sphere containing the disk radius is gradually removed over a 
timescale of 5 Gyr (green line). The initial disk thickness is also
shown for comparison (black line).}
\end{figure}

Although tides and non-axisymmetric instabilities heat the system and remove angular momentum$^{49,50}$, some
residual rotation is present in the dwarf after ten billion years. Figure 6 shows the intrinsic rotational 
velocity as a function of radius; this corresponds to the maximum rotational velocity that can be observed, 
i.e. the velocity measured by observing the dwarf perpendicular to its rotation axis. 
The rotation is negligible near the center but becomes larger than $\sim 3$ km/s at several hundred parsecs from the
center. Interestingly, this is comparable to the rotation inferred for Ursa Minor$^{51,52}$, for which a velocity
gradient around the apparent morphological major axis has been observed out to a distance of a few hundred parsecs.
How much such rotation reflects the original rotation of the dwarf and how much it is the result of dynamical 
evolution driven, for example, by tidal torques, will be subject of a future detailed investigation.

\begin{figure}[t]
\vskip 12.0cm 
{\includegraphics{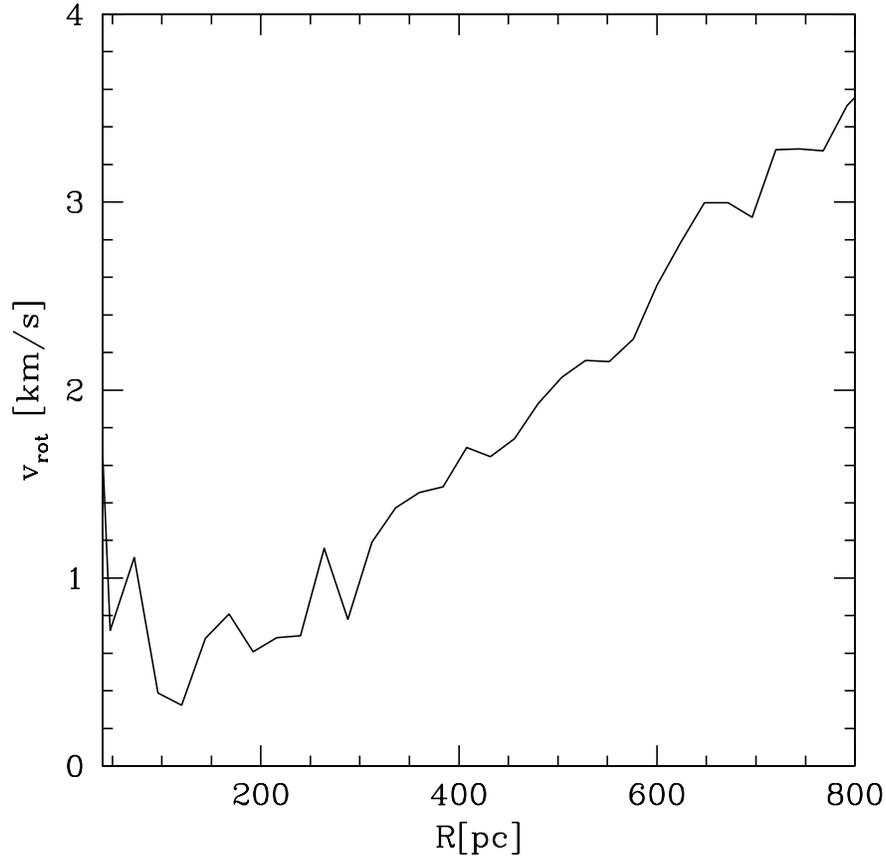}}
\caption[]{\small Rotational velocity profile of the stars as a function of the
distance from the center of the dwarf after 10 Gyr of evolution. 
The profile is azimuthally averaged and spherical radial bins have been used.}
\end{figure}

\subsection{Resolution tests on gas density}

Our simulations lack radiative transfer. Self-shielding of the gas
could reduce the net heating from the cosmic UV background and therefore
reduce the effect of ram pressure. 
Recently, cosmological simulations with SPH hydrodynamics have been used to 
study the formation of dwarf galaxies at high redshift, before as well
as during the reionization epoch$^{53}$.
These include radiative transfer to follow the propagation of UV photons 
produced by a cosmic ionizing background similar to the one employed here and find 
that the gas self-shields efficiently only for densities $\rho > 10^{-22}$ g/cm$^3$.
The gas has always a
density at least an order of magnitude lower than the latter threshold in our
simulations; even during the
phase in which the density increases the most, namely after the bar-driven gas
inflow, we measure $\rho \sim 10^{-23}$ g/cm$^{3}$. 
We conclude that self-shielding
will be negligible given the low density of the gas in the simulated dwarf.
However, resolving the highest gas densities correctly is not trivial in numerical simulations. We tested that
our inference on the importance of self-shielding is not dependent on the adopted resolution
by running the same simulation at three different mass and force resolutions, spanning
a factor of 64 in number of particles (the highest resolution run employs 1.6 million gas particles
and was run only for one orbit). The results are shown in Figure 7. As it
can be seen, between the lowest and the highest resolution case the difference in the
maximum gas density is less than a factor of 2, and it is less than 40\% between the intermediate
and highest resolution cases. The intermediate resolution case corresponds to the standard
resolution adopted in our simulations. The figure also shows that a decrease of a factor of 3
in the gravitational softening implies an increase in the maximum density of only about 30\%
at fixed particle number. 

\begin{figure}[t]
\vskip 11.0cm 
{\includegraphics{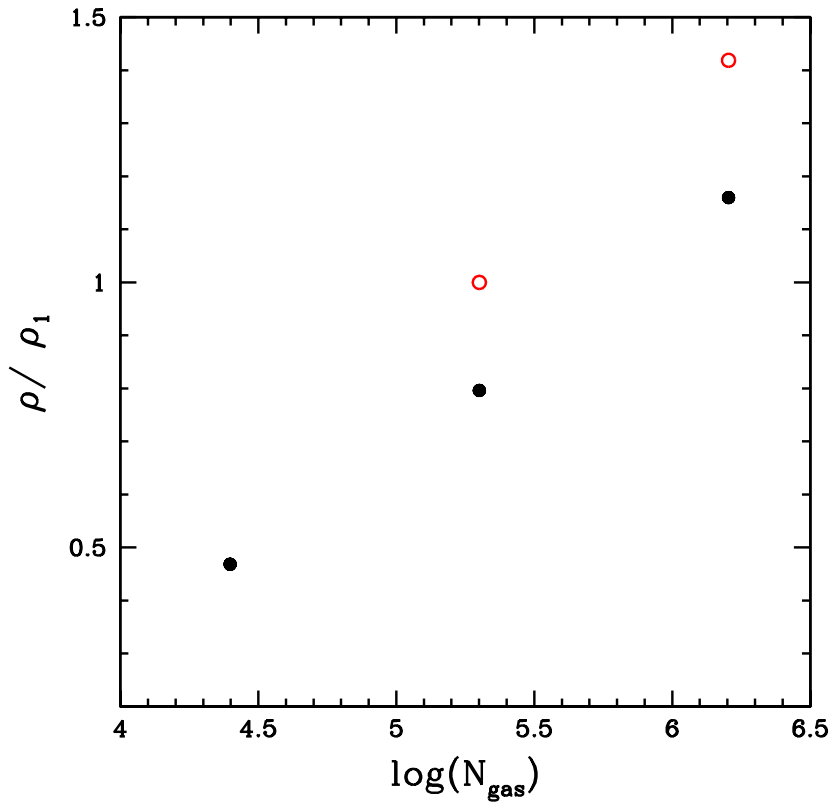}}
\caption[]{\small The maximum density of the gas for runs with different mass and force resolution.
For all the runs the density is measured after bar formation has triggered the gas inflow and its 
value relative to that in the run employed in the Letter is shown 
($\rho_1 = 2\times 10^{-23}$ g/cm$^3$). Red open circles refer to runs in which the gravitational softening is equal to the run discussed in the Letter ($35$ pc), while black filled circles show the results for runs in which the softening was increased by a factor of 3. Therefore, while red and black circles
represent the results of two sequences of runs with varying mass resolution, comparing red and black circles at fixed number of particles shows the effect of changing the gravitational softening.}
\end{figure}

\section{Critical discussion of the results}

\subsection{The various modes of ram pressure stripping}

The gas is first stripped by tides and instantaneous ram pressure as the galaxy approaches pericenter.
This removes the outer disk, outside about two initial disk scale lengths. For such gas,
the ram pressure force, $\rho_{gh_p} {V_p}^2$ ($\rho_{gh_p}$ is the density of the halo near pericenter 
and $V_p$ is the orbital velocity of the galaxy near pericenter, that are, respectively, $10^{-4}$ atoms 
cm$^{-3}$ and 300 km/s) is stronger than the gravitational restoring force, which is mainly provided
by the dark halo. 
More instantaneous ram pressure stripping occurs also at second pericenter as the halo central density,
and hence the gravitational restoring force, has decreased in the meantime due to the tidal
shocks, and the galaxy loses its gas down to about one initial disk scale length. 
Quantitative comparisons with analytic predictions$^{54}$ for instantaneous stripping are difficult 
because tidal forces continuously affect the magnitude
of the gravitational restoring force and participate in the stripping, and the gas is subject
to compressional heating, radiative cooling and an external radiation field.  Nevertheless, it is important to 
notice, that dwarf galaxy models almost identical to the one simulated here have stripping radii within $30\%$  of those predicted 
by the Gunn \& Gott formula$^{54}$ when they are evolved in periodic tube flows that simulate a single
pericenter passage in absence of tidal forces$^{26,42}$. 

However, the gas inside about one disk scale
length lies inside the tidal radius at all times, hence tidal stripping does not play a role; in 
addition, for such gas the gravitational restoring force 
is always higher than the instantaneous ram pressure force. Stripping might still occur
by ablation via Kelvin-Helmoltz instabilities, the so called turbulent stripping, or by laminar 
viscous stripping$^{26,37,55,56,57}$.
As we have also shown elsewhere$^{26}$, Kelvin-Helmoltz instabilities should not occur for dwarfs embedded
in a massive dark halo because of the stabilizing effect of the gravity of the halo. This is the case
even here.
Following existing analytical predictions$^{55}$
we estimate that the critical value of the gravity (in simulation units)
to stabilize our gas disk against Kelvin-Helmoltz is $g_{cr} \sim 0.1$ initially, where $g_{cr} = {2 \pi V^2 / {D R_g}}$
($V \sim 200$ km/s is the average orbital velocity of the galaxy, $R_g = 0.5$ kpc is the scale length
of the gas component, and $D={\rho_g / {\rho_{gh}}}$ is the ratio between the density of the gaseous
disk at $R \sim R_g$ and the time averaged density of gaseous halo along the orbit of the galaxy).
The self-gravity of the dwarf galaxy within $R_g$, which is mainly provided by the dark halo, is instead 
$g  \sim 2.5 > g_{cr}$, where $g=GM_{halo}/{R_g}^2$. We note that $g > g_{cr}$ holds down to 
about 1/5 of the disk scale length, and also for $R > R_g$. Hence the gaseous disk of the dwarf is
initially stable to Kelvin-Helmoltz instabilities. 

When $g > g_{cr}$ the gas can still be stripped
by laminar viscous stripping. The mass loss rate in this case is given by $\dot{M} \sim 12/2.8 \pi 
{R_{g}}^2 V (\lambda_{\rm max} / R_g){(c_1/V)}^{55,57}$, where $\lambda_{\rm max}$
is the longest wavelength not stabilized by gravity. Application of this formula to our dwarf galaxy
in motion through the hot galactic halo yields a characteristic timescale for laminar viscous
stripping which is longer than the Hubble time, mostly because $\lambda_{\rm max} < R_g$. 
However, close to the second pericenter passage, the situation changes (see Figure 8).
The density of the halo within $0.5$ kpc decreases by a factor of 4 as a result of the tidal 
shocks, so that $g$ drops to $\sim 0.6$ at one
disk scale length, now very close to $g_{cr}$ (the latter has also decreased in the meantime, 
$g_{cr} \sim 0.3$, since the mean density of the gas has dropped). 
The gas component of the dwarf is now only marginally stable to Kelvin-Helmoltz instabilities, but using the
formula to calculate the mass loss rate due to Kelvin-Helmoltz$^{55}$ for $R_g = 0.5$ kpc one derives
a characteristic stripping timescale still quite long, i.e. comparable to the Hubble time. Indeed, in
the simulations we do not see Kelvin-Helmoltz instabilities developing before the gas is removed. Instead, assuming that
the longest unstable wavelength is $\lambda_{\rm max} \sim 0.5$ kpc one obtains a laminar viscous timescale 
of $\sim 2$ Gyr. The gas is indeed completely stripped $\sim 1$  Gyr later (it is possible that stripping is
slightly enhanced by artificial viscosity of SPH$^{26}$). 

When the UV background is absent $g >> g_{cr}$ even at the second pericenter passage 
because (1) the density of the halo within 0.5 kpc decreases by only $50\%$ as the cold, concentrated 
gas reduces tidal heating  (see section 5.1 above), and (2) $g_{cr}$ is smaller because the gas is colder 
and denser relative to the 
case where the UV radiation background is included, thereby increasing the value of $D$ (see above) by a factor of 3. 
As a result, the system is very stable to disruption 
by Kelvin-Helmoltz instabilities. Likewise, the longest unstable wavelength is much smaller than the 
radius of the residual gas, which implies a laminar viscous timescale close to the Hubble time, similar
to what we have obtained for the initial conditions. This is consistent 
with our result that the gas is not completely stripped when the UV background is absent.
In conclusion, despite the strongly time-dependent behavior of all the variables involved in the
problem, the results of the simulations are consistent with the expected characteristic timescales
and relative importance of the various ram pressure stripping modes.

\begin{figure}[t]
\vskip 11.6cm 
{\includegraphics{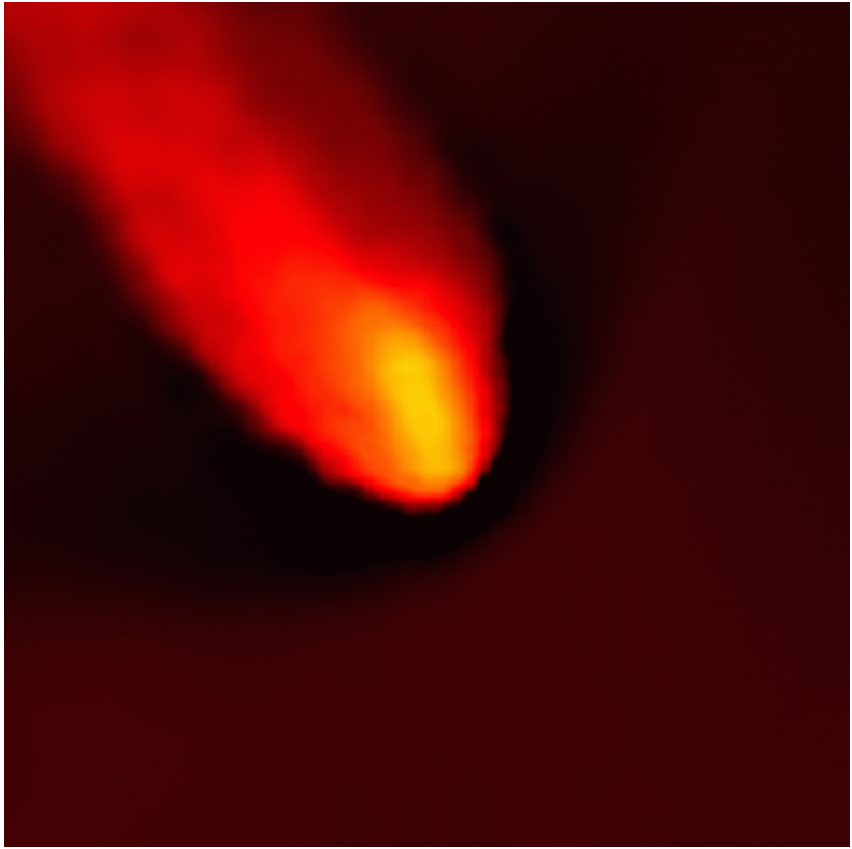}}
\caption[]{\small Color-coded projected density map of the gas component. The box is 3 kpc on a side
and shows the gas distribution view from above the orbital plane. The time is close to the 
second pericenter passage, after about 3 Gyr of evolution. The box corresponds to a zoom in of 
the middle right panel in Figure 1 of the Letter, the density limits are also similar, 
from $10^{-31}$ g/cm$^3$ (black) to $10^{-23}$ g/cm$^3$ (yellow).}
\end{figure}

\subsection{Adding complexity to the scenario}

The ``proximity'' effect due to the local ionizing background from the Milky Way at $z > 1$, not 
included here, could have been even more important than
that of the uniform cosmic UV background.
Galaxy formation simulations that include star formation and the effect of
supernova feedback show that star formation rates comparable to those of
a luminous infrared galaxy (LIRG), $> 10 M_{\odot}$/yr, are attained
around $z \sim 2$ for a Milky Way-type galaxy$^{16}$.
We have run a test with the addition of the local ionizing radiation produced
by the starburst, found complete ram pressure stripping already on the first orbit and, as a 
consequence, an even faster transformation of the stellar component.

The processes discussed in this paper assume that dSphs are bound satellites of a primary galaxy.
Isolated dSphs such as Tucana are found at distances of order twice the expected virial radius of the
Milky Way or M31$^{47,50}$. 
Either they are not bound to either of the primary galaxies or are extreme satellites on
nearly radial orbits (in cosmological simulations a few satellites whose orbital apocenter greatly
exceeds the virial radius are typically found$^{18,47}$). In either case they cannot be objects
that fell early into the halo of a primary galaxy (if bound, their orbital time would be too long
to be consistent with an object infalling at $z > 1$). Instead, their mass-to-light ratios are 
likely dictated by their halo mass rather than by the influence of the environment.
Either their halos have a relatively small mass, $V_{\rm peak} < 20$ km/s, so that reionization alone could have 
been sufficient at suppressing their baryonic content and star formation by inhibiting the collapse of gas 
and/or photoevaporating already accreted gas$^{53}$, or their halos have masses comparable to those of Draco 
but their mass-to-light ratios are much lower, comparable to those of the other dSphs, 
because baryonic sweeping was not effective. Transformation from disk-like progenitors into dSphs
could still be the origin of the diffuse spheroidal structure of Tucana since tidal stirring can 
occur even outside the virial radius of the primary galaxies in cold dark matter models, for example 
between members of subgroups infalling into the primaries$^{58}$. Improving our knowledge on the structure
and properties of isolated dSphs can thus yield important constraints on formation scenarios such as
that proposed in this Letter. Obtaining internal velocities to measure their halo mass and determine
their mass-to-light ratio as well as achieving accurate upper limits on their gas content should be
priorities among the new observational programs of nearby dwarfs.

Finally, we have only considered NFW profiles for the halos of the dwarfs. Recent work$^{45}$ has shown that a cuspy dark matter halo 
fitted by an NFW or Moore profile might develop a core as a result of a strong mass outflow such as that caused by supernovae winds 
after a burst of star formation. Our scenario for baryonic sweeping would be even stronger if dSphs had cored halos.
Indeed, as we have seen in the previous section, the various modes of ram pressure stripping are all 
enhanced if self-gravity is lower, as it would be the case for a cored halo as opposed to a cuspy
halo. Stripping will be much faster especially if the size of the core is comparable to the disk scale 
length since, as we have explained in section 6.1, the inner region is that in which instantaneous stripping 
is not effective and other slower stripping regimes such as viscous or turbulent stripping come into play.
Tidal shocks will also be enhanced because the response of the system will be more impulsive, an effect which is already seen when comparing subhalos 
with NFW profiles having different concentration parameters$^{59}$.
On the other hand, dwarf galaxies having the low gas surface densities and high Toomre parameter of our 
initial model would hardly undergo a strong burst but rather have a fairly low star formation rate$^{35}$,
in which case modifications of the halo profile are unlikely.


\begin{thebibliography}{10}

\bibitem[{Mateo}<1>]{Mateo1998}
{Mateo}, M. {Dwarf Galaxies of the Local Group}. {\it Annual Review
of Astronomy and Astrophysics}. {\bf 36}, 435-506 (1998)

\bibitem[{Grebel}<2>]{Grebel1999}
{Grebel}, E. K. in {\it The Stellar Content of Local Group Galaxies} (eds Whitelock, P. \& Cannon, R.) 17-38, (IAU Symposium 192, ASP, San Francisco 1999)

\bibitem[<3>]{Kleyna02}
Kleyna, J.T., Wilkinson, M.I., Evans, N.W., Gilmore, G. \& Frayn, C. Dark matter in dwarf spheroidals - II.
Observations and modeling of Draco. {\it Mon. Not. R. Astron. Soc.} {\bf 330}, 778-791,
(2002)

\bibitem[<4>]{Chapman}
Chapman, S., Ibata, R., Lewis, G., Fergusom , A., Irwin, M., McConnachie, A. \& Tanvir, N. A Keck DEIMOS Kinematic Study of Andromeda IX: Dark Matter on the Smallest Galactic Scales. {\it Astrophys. J.} {\bf 632}, L87-L90 (2005)


\bibitem[<5>]{Gallagher03}
Gallagher, J. S., Madsen, G. J., Reynolds, R. J., Grebel, E. K. \& Smecker-Hane, T. A. {A Search for Ionized Gas in the Draco and Ursa Minor Dwarf Spheroidal Galaxies}. {\it Astrophys. J}. {\bf 588}, 326-330 (2003).

\bibitem[<6>]{Kaza2004}
Kazantzidis, S., Mayer, L., Mastropietro, C., Diemand, J., Stadel, J. \& Moore, B. {Density 
profiles of Cold Dark Matter substructure: implications for the missing-satellites problem}. {\it Astrophys. J.} {\bf 608}, 663-679 (2004)

\bibitem[<7>]{Lin83} Lin, D.N.C. \& Faber, S.M. Some implications of nonluminous matter in dwarf spheroidal galaxies. {\it Astrophys. J.} {\bf 266}, L21-L25 (1983)


\bibitem[<8>]{Dekel86}
Dekel, A. \& Silk, J. {The origin of dwarf galaxies, cold dark matter, and biased galaxy for
mation}. {\it Astrophys. J.} {\bf 303}, 39-55 (1986)

\bibitem[<9>]{Bullock01}
Bullock, J., Kravtsov, A. \& Weinberg, D.H. {Reionization and the abundance of galactic satellites}. {\it Astrophys. J}. {\bf 539}, 517-521 (2000)

\bibitem[<10>]{Mayer01}
Mayer, L., Governato, F., Colpi, M., Moore, B., Quinn, T., Wadsley, J., Stadel, J. \& Lake, G. {The Metamorphosis of Tidally Stirred Dwarf Galaxies}. {\it Astrophys. J.} {\bf 559}, 754-784 (2001)
 

\bibitem[<11>]{Blok97}
de Blok, W.J.G. \& McGaugh, S.S. {The dark and visible matter content of low surface  brightness disc galaxies.} {\it Mon. Not. R. Astron. Soc.} {\bf 290}, 533-552 (1997)

\bibitem[<12>]{} Mayer, L. \& Moore, B, The baryonic mass-velocity relation: clues to feedback processes during structure formation and the cosmic baryon inventory. {\it Mon. Not. R. Astron. Soc.} {\bf 354}, 477-484 (2004)

\bibitem[<13>]{Susa}
Susa, H. \& Umemura, M. Formation of Dwarf Galaxies during the Cosmic Reionization. {\it Astrophys. J.} {\bf 600}, 1-16 (2004)

\bibitem[<14>]{Mac99}
Mac Low, M.M. \& Ferrara, A. {Starburst-driven Mass Loss from 
Dwarf Galaxies: Efficiency and Metal Ejection}. {\it Astrophys. J}. {\bf 513}, 142-155 (1999)

\bibitem[<15>]{Einasto74}Einasto, J., Kaasik, A. \& Saar, E. {Dynamic evidence on massive coronas 
of galaxies}. {\it Nature}. {\bf 252}, 111-113 (1974)

\bibitem[<16>]{Mar03}
Marcolini, A., Brighenti, F. \& D'Ercole, A. Three-dimensional simulations of the interstellar medium
in dwarf galaxies - I. Ram pressure stripping. {\it Mon. Not. R. Astron. Soc.} {\bf 345}, 1329-1339 (2003)

\bibitem[<17>]{Mori02} Mori, M. \& Burkert, A. Gas stripping of dwarf galaxies in clusters of galaxies,
{\it Astrophys. J.} {\bf 538}, 559-568 (2000)

\bibitem[<18>]{Krav04}
Kravtsov, A. V., Gnedin, O. Y.\& Klypin, A. A. {The Tumultuous Lives of Galactic Dwarfs and 
the Missing Satellites Problem}. {\it Astrophys. J}. {\bf 609}, 482-497 (2004)


\bibitem[<19>]{Diem05}
Diemand, J., Madau, P. \& Moore. B. {The distribution and kinematics of early high-sigma peaks in present-day haloes: implications for rare 
objects and old stellar populations}.  {\it Mon. Not. R. Astron. Soc.} {\bf 364}, 367-383 (2005)

\bibitem[<20>]{Gov04}
Governato, F., Mayer, L., Wadsley, J., Gardner, J. P., Willman, Beth, Hayashi, E., Quinn, T., Stadel, J.\& Lake, G. {The Formation of a Realistic Disk Galaxy in Lambda-dominated Cosmologies}. {\it Astrophys. J.} {\bf 607}, 688-696, 2004

\bibitem[<21>]{Verde02}
Verde, L., Oh, P.S. \& Jimenez, R. {The abundance of dark galaxies}.{\it Mon. Not. R. Astron. Soc.} {\bf 336}, 541-549 (2002)

\bibitem[<22>]{Haard96}
Haardt, F. \& Madau, P. {Radiative Transfer in a Clumpy 
Universe. II. The Ultraviolet Extragalactic Background}. {\it Astrophys. J}. {\bf 461}, 20-37 (1996)

\bibitem[<23>]{Gunn72}
Gunn, J. E. \& Gott, J. R. I. {On the Infall of Matter  Into Clusters of Galaxies and Some Effects on Their Evolution}. {\it Astrophys. J}. {\bf 176}, 1-19 (1972)

\bibitem[<24>] {Munoz05}
Munoz, R.P. {\it et al}. {Exploring Halo Substructure with Giant Stars: The Velocity Dispersion Profiles of the Ursa Minor and 
Draco Dwarf Spheroidal Galaxies at Large Angular Separations}. {\it Astrophys. J}. {\bf 631}, L137-L141 (2005)


\bibitem[<25>]{} Willman, B., Governato, F. Dalcanton, J.J., Reed, D. \& Quinn, T. The observed and
predicted spatial distribution of Milky Way satellites. {\it Mon. Not. R. Astron. Soc.} {\bf 353}, 639-646 (2004)


\bibitem[<26>] {Moore99}
Moore, B., Ghigna, S., Governato, F., Lake, G., Quinn, T., Stadel, J. \& Tozzi, P. {Dark Matter Substructure within Galactic Halos}. {\it Astrophys. J}. {\bf 524}, L19-L22 (1999)

\bibitem[<27>] {Willman05}
Willman, B. {\it et al.} {A New Milky Way Dwarf Galaxy in Ursa  Major}. {\it Astrophys. J}. {\bf 626}, L85-L88 (2005)

\bibitem[<28>]{Kleyna05}
Kleyna, J.T., Wilkinson, M.I., Evans, Wyn N., \& Gilmore, G. {Ursa Major:a missing low mass CDM halo?}. {\it Astrophys. J.} {\bf 630}, L141-L144
(2005)

\bibitem[<29>]{}Walker, M .G., Mateo, M., Olszewski, E.W., Bernstein, R., Wang, X. \& Wooodrofe, M., Internal Kinematics of the Fornax Dwarf Spheroidal Galaxy, {\it Astron. J}. {\bf 131}, 2114-2139 (2006)

\end{thebibliography}

\begin{thebibliography}{}

\bibitem[<1>]{}
Wadsley, J., Stadel, J., \& Quinn, T., Gasoline: a flexible, parallel implementation of TreeSPH. {\it New Astr.} {\bf 9}, 137-158 (2004)

\bibitem[<2>]{}
Barnes, J., \& Hut, P., A Hierarchical O(NlogN) Force-Calculation 
Algorithm. {\it Nature} {\bf 324}, 446-449 (1986)

\bibitem[<3>]{}
Hernquist, L., Bouchet, F., and Suto, Y. , Application of the Ewald 
method to cosmological N-body simulations. {\it Astrophys. J. Supp} {\bf 75}, 
231-240 (1991)

\bibitem[<4>]{}
Gingold, R.A. \& Monaghan, J.J. Smoothed particle hydrodynamics: theory 
and application to non-spherical stars. {\it Mon. Not. R. Astron. Soc.}
{\bf 181}, 375-389 (1977)

\bibitem[<5>]{}
Monaghan, J.J., Smoothed particle hydrodynamics. {\it Annual. Rev. Astron, Astrophys.} {\bf 30}, 543-574 (1992)


\bibitem[<6>]{} Springel, V., \& Hernquist, L.
Cosmological smoothed particle hydrodynamics simulations: the entropy equation. {\it Mon. Not. R. Astron. Soc.} {\bf 333}, 649-664 (2002)

\bibitem[<7>]{}
Haardt , F. \& Madau. P., Radiative Transfer in a Clumpy Universe. II. 
The Ultraviolet Extragalactic Background. {\it Astrophys. J.} {\bf 461}, 20-37(1996)

\bibitem[<8>]{}
Alvarez, M.A., Shapiro, P., Kyungjin A. \& Iliev, I.T. Implications of WMAP 3 Year Data for Sources
of Reionization. {\it Astrophys. J.} {\bf 644} L101-L104 (2006)

\bibitem[<9>]{} Grebel, E.K., Gallagher, J.S., \& Herbeck, D. The Progenitors of Dwarf Spheroidal Galaxies. {\it Astron. J.} {\bf 125}, 
1926-1939 (2003)


\bibitem[<10>]{}
Balsara, D.S., von Neumann stability analysis of smoothed particle 
hydrodynamics:suggestions for optimal algorithms. {\it J.Comput. Phys.} {\bf 121} 357-372 (1995)


\bibitem[<11>]{Gov04}
Governato, F., Mayer, L., Wadsley, J., Gardner, J. P., Willman, Beth, 
Hayashi, E., Quinn, T., Stadel, J.\& Lake, G, {The Formation of a 
Realistic Disk Galaxy in Lambda-dominated Cosmologies}. {\it Astrophys. J.} {\bf 607}, 688 (2004)

\bibitem[<12>]{}
{Klypin} A.,  {Zhao} H.,    {Somerville} R.~S., $\Lambda$ CDM-based Models for the Milky Way and M31. I. Dynamical Models, Astrophys. J., {\bf 573}, 597-613 (2002)

\bibitem[<13>]{}
Efstathiou, G., {\it et al.},Evidence for a non-zero Lambda and a low 
matter density from a combined analysis of the 2dF Galaxy Redshift 
Survey and cosmic microwave background anisotropies. {\it Mon. Not. R. Astron. Soc.} {\bf 330}, 
L29-L35 (2002)

\bibitem[<14>]{}
Power, C., {\it et al.},The inner structure of LambdaCDM haloes - I. A 
numerical convergence study. {\it Mon. Not. R. Astron. Soc.} {\bf 338}, 
14-34 (2003)

\bibitem[<15>]{}
{Gardner} J.~P., Dependence of Halo Properties on Interaction History, 
Environment, and Cosmology, {\it Astrophys. J.}, {\bf 557}, 616-625 (2001)


\bibitem[<16>]{}
Governato {\it et al.} {\it Mon. Not. R. Astron. Soc.} in press


\bibitem[<17>]{}
Katz, N., Quinn, T., Bertschinger, E., \& Gelb, J.M., Formation of Quasars 
at High Redshift,  {\it Mon. Not. R. Astron. Soc.} {\bf 270}, L71-L74 (1994)

\bibitem[<18>]{}
Ghigna, S., Moore, B., Governato, F., Lake, G., Quinn, T., \& Stadel, J., 
Dark matter haloes within clusters, {\it Mon. Not. R. Astron. Soc.} {\bf 300}, 146-162 (1998)


\bibitem[<19>]{}
Moore, B., {\it et al.}, Cold collapse and the core catastrophe, {\it Mon. Not. R. astron. Soc.} {\bf 310}, 1147-1152 (1999)
  
\bibitem[<20>]{}
Mastropietro, C., Moore, B., Mayer, L., Debattista, V., Piffaretti, R., 
\& Stadel, J., Morphological evolution of discs in clusters.  {\it Mon. Not. R. astron. Soc.} {\bf 364}, 
607-619 (2005)


\bibitem[<21>]{}
Hernquist, L., N-body realizations of compound galaxies, {\it Astrophys. J. Supp.} {\bf 86}, 389-400 (1993)


\bibitem[<22>]{}
Springel, V. \& White, S.D.M.,Tidal tails in cold dark matter cosmologies. {\it Mon. Not. R. Astron. Soc.} {\bf 307}, 162-178 (1999)


\bibitem[<23>]{}
{Fall} S.~M., \& {Efstathiou} G.,  Formation and rotation of disc galaxies with haloes,  {\it Mon. Not. R. Astron. Soc.} {\bf 193}, 189-206 (1980)
 
\bibitem[<24>]{} Mo, H. J., Mao \& White, S. D. M. , The formation of galactic discs,  {\it Mon. Not. R. Astron. Soc.} {\bf 295}, 319-336 (1998)

\bibitem[<25>]{}Mayer, L., {Moore}, B., {Quinn}, T., {Governato}, F., \& 
{Stadel}, J., Tidal debris of dwarf spheroidals as a probe of structure 
formation models, {\it Mon. Not. R. Astron. Soc.} {\bf 336}, 119-130  (2002)

\bibitem[<26>]{} Mayer, L., Mastropietro, C., Wadsley, J., Stadel, J., 
\& Moore, B., Simultaneous ram pressure and tidal stripping; how dwarf 
spheroidals lost their gas, {\it Mon. Not. R. Astron. Soc.} {\bf 369}, 
1021-1038 (2006)

\bibitem[<27>]{}
Colin, P., Klypin, A., Valenzuela, O., \& Gottloeber, S., Dwarf Dark 
Matter Halos. {\it Astrophys. J.} {\bf 612}, 50-57 (2004)

\bibitem[<28>]{}
Navarro, J. F., Frenk, C. S., White, S. D. M., The Structure of Cold Dark Matter Halos, 
{\it Astrophys. J.} {\bf 462}, 563-575 (1996)

\bibitem[<29>]{Jim03}
Jimenez, R., Verde, L., \& Oh, S. P., Dark halo properties from 
rotation curves, {\it Mon. Not. R. Astron. Soc.} {\bf 339},  243-259 (2003)

\bibitem[<30>]{} Mayer, L., \& Moore, B, The baryonic mass-velocity 
relation: clues to feedback processes during structure formation and the 
cosmic baryon inventory, {\it Mon. Not. R. Astron. Soc.} {\bf 354}, 477-484 (2004)

\bibitem[<31>]{Verde02}
Verde, L., Oh, S.P., \& Jimenez, R., {The abundance of dark galaxies}, 
{\it Mon. Not. R. Astron. Soc.} {\bf 336}, 541-549 (2002)


\bibitem[<32>]{}
Wong. T., \& Blitz, L., 2002, The Relationship between Gas Content and 
Star Formation in Molecule-rich Spiral Galaxies, {\it Astrophys. J.} {\bf 569}, 157-183 (2002)

\bibitem[<33>]{}
Vallenari, A., Schmidtobreick, L., \& Bomans, D.J., The star formation 
history of the LSB galaxy UGC 5889, {\it Astr. and Astrophys.} {\bf 435}, 
821-829 (2005)

\bibitem[<34>]{}
Schaye, J., Star Formation Thresholds and Galaxy Edges: Why and Where. {\it Astrophys. J.} {\bf 609}, 667-682 (2004)

\bibitem[<35>]{}
Li, Y., Mac Low, M., \& Klessen, R., Star Formation in Isolated Disk Galaxies. I. Models and Characteristics of Nonlinear Gravitational Collapse, {\it Astrophys.J.} {\bf 626}, 823-843, (2005)

\bibitem[<36>]{}
Klein, R., McKee, C.F., \& Colella, P., On the hydrodynamic interaction of shock waves with interstellar clouds. 1: Nonradiative shocks in small clouds, {\it Astrophys. J.} {\bf 420}, 213-236 (1994)

\bibitem[<37>]{}
Mac Low, M, \& Zahnle, K., Explosion of comet Shoemaker-Levy 9 on entry into the Jovian atmosphere. {\it Astrophys. J.} {\bf 434}, L33-L36 (1994)


\bibitem[<38>]{}
Bate, M., \& Burkert, A., Resolution requirements for smoothed particle hydrodynamics calculations 
with self-gravity,  {\it Mon. Not. R. Astron. Soc.} {\bf 288}, 1060-1072 (1997)

\bibitem[<39>]{}
Dehnen, W., \& Binney, J., Mass models of the Milky Way, {\it Mon. Not. R. Astron. Soc.} {\bf 294}, 429-434 (1998)

\bibitem[<40>]{Li95}
Lin, D.N.C., Jones, B.F., \& Klemola, A.R. , The motion of the 
Magellanic clouds, origin of the Magellanic Stream, and the mass of the 
Milky Way {\it Astrophys. J.} {\bf 439}, 652-671 (1995)


\bibitem[<41>]{}
Sembach, K.R., {\it et al.}, Highly Ionized High-Velocity Gas in the Vicinity 
of the Galaxy, {\it Astrophys. J. Supp.} {\bf 146}, 165-208 (2003)


\bibitem[<42>]{}
Mastropietro, C., Moore, B, Mayer, L., Wadsley, J., \& Stadel, J., The gravitational and hydrodynamical interaction between the Large 
Magellanic Cloud and the Galaxy, Mon. Not. R. Astron. Soc., {\bf 363}, 509-520 (2006)

\bibitem[<43>]{}
Navarro, J. F., Eke, V.R., \& Frenk, C.S., The cores of dwarf galaxy 
haloes, {\it Mon. Not. R. Astron. Soc.} {\bf 283}, L72-L78 (1996)

\bibitem[<44>]{}
Gnedin, O., \& Zhao, H., Maximum feedback and dark matter profiles of dwarf galaxies {\it Mon. Not. R. Astron. Soc.} {\bf 333}, 299-306 (2002)


\bibitem[<45>]{} Read, J, \& Gilmore, G., Mass loss from dwarf spheroidal galaxies: the origins of shallow dark matter cores and exponential surface brightness profiles, {\it Mon. Not. R. Astron. Soc.} {\bf 356}, 107-124 (2005)

\bibitem[<46>]{}
Dubinski, J., \& Carlberg, R.G., The structure of cold dark matter halos, {\it Astrophys. J.} {\bf 378}, 496-503 (1991)

\bibitem[<47>]{}
{Mateo}, M., 1998, {Dwarf Galaxies of the Local Group}, {\it Annual Review of Astronomy and Astrophysics} {\bf 36}, 435-506 (1998)

\bibitem[<48>]{}
James, P., An infrared study of dwarf galaxies in the Virgo cluster, {\it Mon. Not. R. Astron. Soc.} {\bf 250}, 544-554 (1991)

\bibitem[<49>]{Ma01a}
Mayer, L., Governato, F., Colpi, M., Moore, B., Quinn, T., Wadsley, J.,
Stadel, J., Lake G., Tidal Stirring and the Origin of Dwarf Spheroidals in the Local Group, {\it Astrophys. J.} {\bf 547}, L123-L127 (2001)

\bibitem[<50>]{Mayer01}
Mayer, L., Governato, F., Colpi, M., Moore, B., Quinn, T., Wadsley, J., Stadel, J., \& Lake, G., {The Metamorphosis of Tidally Stirred Dwarf Galaxies}, {\it Astrophys. J.} {\bf  559}, 754-784 (2001)


\bibitem[<51>]{}
Hargreaves, J. C., Gilmore, G, Irwin, M. J.\& Carter, D., A Dynamical Study of the Ursa-Minor Dwarf Spheroidal Galaxy,  {\it Mon. Not. R. Astron. Soc.} {\bf 271}, 693-705 (1994)

\bibitem[<52>]{}    
Armandroff, T. E., Olszewski, E. W, Pryor, C., The Mass-To-Light Ratios of the Draco and Ursa Minor Dwarf Spheroidal Galaxies.I. Radial Velocities from Multifiber Spectroscopy, {\it Astron. J.} {\bf 110}, 2131-2165 (1995)

\bibitem[<53>]{Susa}
Susa, H., \& Umemura, M., Formation of Dwarf Galaxies during the Cosmic Reionization, {\it Astrophys. J.} {\bf 600}, 1-16 (2004)

\bibitem[<54>]{GG}
Gunn, J. E., Gott, J. R. I., {On the Infall of Matter Into Clusters of Galaxies and Some Effects on 
Their Evolution}, {\it Astrophys. J} {\bf 176}, 1-19 (1972)

\bibitem[<55>]{}
Murray, S. D., White, S. D. M., Blondin, J. M., \& Lin, D. N. C., Dynamical instabilities in two-phase media and the minimum masses of stellar systems,  {\it Astrophys. J.} {\bf 407}, 588-596 (1993)

\bibitem[<56>]{}
Quilis, V., Moore, B \& Bower, R., Gone with the Wind: The Origin of S0 Galaxies in Clusters, {\it Science.} {\bf 288}, 1617-1620 (2000)

\bibitem[<57>]{}
Nulsen, P. E. J., Transport processes and the stripping of cluster galaxies, {\it Mon. Not. R. Astron. Soc.} {\bf 198}, 1007-1016 (1982)

\bibitem[<58>]{}
Kravtsov, A. V., Gnedin, O. Y., Klypin, A. A. {The Tumultuous Lives of Galactic Dwarfs and the Missing Satellites Problem}, {\it Astrophys. J} , {\bf 609}, 482-497 (2004)

\bibitem[<59>]{}
Kazantzidis, S., Mayer, L., Mastropietro, C., Diemand, J., Stadel, J., \& Moore, B., {Density 
profiles of Cold Dark Matter substructure: implications for the missing-satellites problem}, {\it Astrophys. J.}, {\bf 608}, 663-679 (2004)


\end{thebibliography}
\end{document}